\documentclass[pra,twocolumn,preprintnumbers,amsmath,amssymb,superscriptaddress,showpacs,longbibliography]{revtex4-1}

\usepackage{graphicx}
\usepackage{amsmath}
\usepackage{graphics}
\usepackage{amssymb}
\usepackage{amsfonts,epsfig}
\usepackage{dsfont}
\usepackage{natbib}
\usepackage{enumitem}
\usepackage{physics}
\usepackage{color}
\usepackage[mathscr]{eucal}
\usepackage{ulem}

\allowdisplaybreaks

\newcommand{ \sinc }{ \mathrm{sinc} }						
\newcommand{ \csp }[1]{v_{#1}}								
														
\begin{document}

\title{The dynamical-decoupling-based spatiotemporal noise spectroscopy}
\author{Jan Krzywda}
\affiliation{Institute of Physics, Polish Academy of Sciences, al.~Lotnik{\'o}w 32/46, PL 02-668 Warsaw, Poland}
\author{Piotr Sza\'{n}kowski}\email{piotr.szankowski@ifpan.edu.pl}
\affiliation{Institute of Physics, Polish Academy of Sciences, al.~Lotnik{\'o}w 32/46, PL 02-668 Warsaw, Poland}
\author{{\L}ukasz Cywi\'{n}ski}\email{lcyw@ifpan.edu.pl}
\affiliation{Institute of Physics, Polish Academy of Sciences, al.~Lotnik{\'o}w 32/46, PL 02-668 Warsaw, Poland}

\begin{abstract} Here we demonstrate how the standard, temporal-only, dynamical-decoupling-based noise spectroscopy method can be extended to also encompass the spatial degree of freedom. This {\it spatiotemporal} spectroscopy utilizes a system of multiple qubits arranged in a line that are undergoing pure dephasing due to environmental noise. When the qubits are driven by appropriately coordinated sequences of $\pi$ pulses the multi-qubit register becomes decoupled from all components of the noise, except for those characterized by frequencies and wavelengths specified by the pulse sequences. This allows for employment of the procedure for reconstruction of the two-dimensional spectral density that quantifies the power distribution among spatial and temporal harmonic components of the noise. 
\end{abstract}

\date{\today}
		
\maketitle

\section{Introduction}
	
Single qubits have been successfully used to characterize the frequency spectrum of environmental fluctuations affecting their coherence \cite{Degen_RMP17,Szankowski_JPCM17}. The basic principle is to subject the qubit to a periodic perturbation---a sequence of unitary operations or measurements---in order to make it sensitive only to certain frequencies of these fluctuations \cite{Kofman_PRL01,Kofman_PRL04,Gordon_PRL08,Alvarez_PRL11,Bylander_NP11,Yuge_PRL11,Kotler_Nature11}.
Currently the most commonly used method of turning a qubit into such a frequency-discriminating sensor is to drive it with a periodic sequence of pulses that cause an effectively instantaneous $\pi$ rotations of the qubit's Bloch vector. Such a sequence of rotations acts as a {\it filter} of environmental noise \cite{Cywinski_PRB08,Biercuk_Nature09,Biercuk_JPB11}: low frequency noise is strongly suppressed, while at discrete frequencies determined by the pulse spacing the filter has both notches, completely suppressing noise at given frequencies \cite{Malinowski_NN17}, and narrow passbands that single out frequencies of noise that are allowed to have dominant influence on qubit's dephasing \cite{Alvarez_PRL11,Bylander_NP11,Yuge_PRL11,Kotler_Nature11}.
The application of such a sequence leads then to dynamical decoupling \cite{Viola_PRA98,Uhrig_PRL07,Gordon_PRL08,Khodjasteh_NC13,Suter_RMP16} that on one hand extends qubits's coherence time by suppressing the influence of most of environmental fluctuations, and on the other makes the qubit sensitive to noise frequencies determined by the inverse of applied sequence period \cite{Gordon_PRL08,Alvarez_PRL11,Bylander_NP11,Yuge_PRL11,Kotler_Nature11}.
This method has been successfully implemented with many kinds of qubits: superconducting circuits \cite{Bylander_NP11}, trapped ions \cite{Kotler_Nature11}, ultracold atoms \cite{Almog_JPB11}, semiconductor-based quantum dots \cite{Medford_PRL12,Malinowski_PRL17,Chan_PRAPL18}, donors in silicon \cite{Muhonen_NN14}, and nitrogen-vacancy centers in diamond \cite{Staudacher_Science13,Romach_PRL15}. The information obtained in this way led to multiple new insights into the dynamics of environments affecting various kind of qubits.

Such a wide experimental adoption of single-qubit noise spectroscopy method, and recent progress in experimental control over multi-qubit registers realized in various platforms \cite{Monz_PRL11,Barends_Nature14,Nigg_Science14,Ballance_PRL16,Song_PRL17,Zajac_PRAPL16,Fujita_NPJQI17,Vandersypen_NPJQI17,Malinowski_arXiv18,Klimov_PRL18,Neill_Science18,Friis_PRX18}, 
 has motivated recent theoretical research into possibility of using multi-qubit registers to obtain more detailed information about their environment. For example, in \cite{Szankowski_PRA16,Paz_PRA17} the authors considered multiple qubits that, although coupled to shared environmental degrees of freedom, were described as being exposed to local---and possibly correlated---noises representing fluctuations at the physical location of each qubit. Then, with an application of properly coordinated sequences of pulses one can perform spectroscopy of these local fluctuations, but more importantly, also of the cross-correlations between noises affecting different qubits. The special cases of perfectly correlated local noises (all cross-correlations equal to local auto-correlations), and of completely independent noises (vanishing cross-correlations), were considered already in first papers devoted to decoherence of multi-qubit states \cite{Palma_PRSLA96,Duan_PRA98}. Using multiple entangled qubit to enhance the precision of spectroscopy of perfectly correlated noises was discussed \cite{Muller_SR18}. However, quantitative investigations of more general forms of cross-correlations of multiple classical \cite{Szankowski_PRA16,Paz_NJP16} and quantum noises \cite{Krzywda_SR16,Paz_NJP16,Paz_PRA17,Kwiatkowski_PRB18} (the term ``quantum noise'' indicates a case when the interaction between the environment and qubits requires full quantum mechanical treatment as it cannot be approximated with an exposition to effective external stochastic process), have been subject to closer investigation only quite recently. Such investigations are currently of high importance for ongoing and near-future development of multi-qubit quantum registers for various applications. The cross-correlations of local noises are known to have significant influence on quantum metrological applications of systems of multiple entangled qubits \cite{Kolodynski_NJP13,Jeske_NJP14,Beaudoin_PRA18,Haase_arXiv18,Arrad_PRL14,Kessler_PRL14,Demkowicz_PRX17,Zhou_NC18}, and quantum error correction \cite{Devitt_RPP13,Brown_RMP16}---a central issue for long-term prospects of quantum computation---that crucially relies on assumptions about correlations in decoherence processes of multiple qubits \cite{Ng_PRA09,Novais_PRL13,Hutter_PRA14,Preskill_QIC13,Bermudez_PRX17,Bermudez_arXiv18}.

However, such a  framework for description of noise probed by multi-qubit register has its drawbacks. The main one is that the local noise is defined as temporal fluctuations that are experienced by a concrete qubit placed in a given spatial location. Therefore, when the register is modified---e.g., qubits are relocated, added or removed---the set of local noises defined with respect to the original register configuration ceases to exist and is replaced by a new set that characterizes the decoherence of the modified register. Moreover, when no additional information or assumptions regarding the structure and nature of the environment are brought in from outside (e.g., see Ref.~\cite{Krzywda_PRA17}), this framework in itself does not offer any means of relating the properties of local noises affecting qubits in their new configuration with the properties of noises corresponding to the previous, or any other, configuration. Intuitively this looks like a failure of the approach as it is natural to presume that the local noises are ultimately caused by the environment common among all the qubits no matter their current configuration, and thus they should be related in a way that reflects it, and the means for exhibiting those relations should be somehow built-in into the used framework. 

Here we propose a subtle but consequential evolution in the description of noise probed by the multi-qubit register that have a potential to satisfy these expectations: instead of a set of local noises attached to qubits we consider a {\it noise field} that depends on both time and position. Then, the noise field evaluated at a given spatial argument is identical to local noise attached to a qubit that would be placed at this location, and the cross-correlations between local noises are equivalent to auto-correlations of the field with fixed values of spatial arguments corresponding to positions of qubits that would define these noises. The special cases of independent and perfectly correlated local noises observed in a given configuration of the register are explained by the spatial range of the noise field's auto-correlation, the so-called correlation length; the former case occurs when the range is much shorter than the inter qubit distance, while the latter case is realized when it is much longer than the length of the register. Thus, these two extreme cases are analogous to the white noise and quasi-static noise limits of temporal fluctuations. A straightforward conclusion from this discussion is that the noise field's auto-correlation in its entirety is a much richer source of information about the environmental noise than any finite set of cross-correlations of local noises. The question thus becomes: Is it possible to access this information, and if it is, how can it be accomplished? The main purpose of this paper is to provide a positive answer to this question and to present a setup in which the measurement of multi-qubit coherences is related in possibly most direct way with the field's auto-correlation, or rather with its two-dimensional Fourier transform---the {\it spatiotemporal spectral density} of the noise field.

The design of the setup is derived from solutions developed for previously considered multi-qubit spectrometers of local noises and their cross-correlations. The key element is the ability to filter particular noise frequencies together with the {\it wavevectors} (or wavelengths) of spatial fluctuations; in our setup it is achieved with coordination of pulse timings between qubits that is specific to given spatial configuration of the register. While this design of spatiotemporal frequency filter is broadly applicable to all noise fields, the detailed scheme for extracting the data on the spectral density from the multi-qubit coherence measurements that we present here works for stationary and spatially uniform fields having Gaussian statistics (i.e., fields completely specified by their average value and the auto-correlation).

The paper is organized in the following way. First, in Section \ref{sec:intro} we give an overview that explains the principles of operation of the method in an intuitive way. Then, in Section \ref{sec:tech} we present all the necessary formal derivations showing that our setup indeed acts as a spatiotemporal frequency filter of the noise field. In Sec.~\ref{sec:dist_proc} we discuss the contribution to decoherence that contains the spectroscopic information about the noise, and we explain how to reliably extract this contribution from the measured data. There we also develop the supporting methods that allow to deal with expected difficulties that would not be an issue for the temporal-only variant of spectroscopy, but here can make or break the whole method. This section is rather technical and can be skipped in the first reading, or if the reader is more interested in the connection between coherence of qubits and spatiotemporal correlations of the noise field, and less in details of how this connection can be practically exploited. In Sec.~\ref{sec:examples} we present an example of spectral density reconstruction in a numerically simulated experiment. Finally, in Sec.~\ref{sec:end} we discuss and summarize the results.

\section{Overview}\label{sec:intro}

\begin{figure}[t]
\centering
\includegraphics[width=\columnwidth]{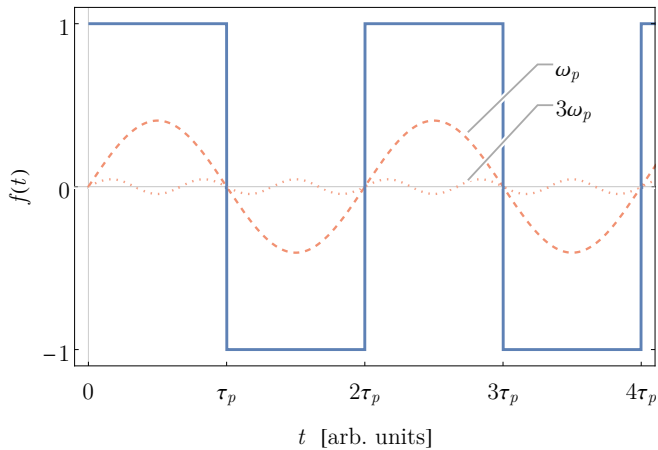}
\caption{The time-domain filter function $f(t)$ (solid blue line). The moments when the function switches sign ($t=\tau_p,2\tau_p,3\tau_p\ldots$) coincide with the timings of $\pi$-pulse applications. The first two harmonic components of $f(t)$ oscillating with frequencies $\omega_p=\pi/\tau_p$ and $3\omega_p$ are depicted in orange dashed and dotted lines. The amplitude of a harmonic is set to be equal to the modulus square of the corresponding Fourier series coefficient calculated in respect to a single period of $f(t)$: $(2\tau_p)^{-1}\int_0^{2\tau_p}dt f(t)e^{-i m\omega_p t}$, with $m=1,3$.}
\label{fig:f_harmonics}
\end{figure}

Consider a qubit localized at a point in space $\mathbf{r}_\mathrm{Q}$, with energy splitting $\hat H_\mathrm{1\mathrm{Q}}=\Omega\hat{\sigma}_z/2$, where $\hat{\sigma}_z$ is the $z$-component of qubit's Pauli operator vector. It is coupled to the environmental noise via phase Hamiltonian $\hat V_{1\mathrm{Q}} = \xi(\mathbf{r}_\mathrm{Q},t)\hat\sigma_z/2$, where the stochastic function $\xi(\mathbf{r},t)$ describes the noise field emitted by the environment. During the evolution the qubit is subject to a control sequence where $\pi$-pulses are applied cyclically, with fixed interval $\tau_p$ between the pulses. The resultant dynamics affect only the off-diagonal elements (in basis of eigenstates $\hat\sigma_z|{\pm}\rangle = {\pm}|{\pm}\rangle$) of qubits' density matrix $\hat\varrho_\mathrm{Q}$. The effect of such a driving is most clearly captured using Heisenberg picture of ladder operator $\hat\sigma_+ = (\hat\sigma_x+i\hat\sigma_y)/2$, for which $\langle +|\hat\varrho_\mathrm{Q}(T)|{-}\rangle \propto \mathrm{Tr}[ \hat\sigma_{+}(T)\hat\varrho_\mathrm{Q}(0) ]$:
\begin{align}
\nonumber
\hat{\sigma}_+(T) &= \overline{e^{i\frac{1}{2}\hat{\sigma}_z\phi_\xi(T)}\hat{\sigma}_{+}e^{-i\frac{1}{2}\hat{\sigma}_z\phi_\xi(T)}}\\
&=	\overline{\exp[-i \phi_\xi(T)]}\,\hat{\sigma}_{+} \, ,
\end{align}
where $T$ is the duration of the evolution, and the symbol $\overline{(\ldots)}$ represents the averaging over realizations of the stochastic process $\xi$. The stochastic phase accumulated for a single trajectory of the noise is given by
\begin{equation}\label{eq:1_qubit_phase}
\phi_\xi(T) = \int_0^T dt f(t)\xi(\mathbf{r}_\mathrm{Q},t)\,,
\end{equation}
where $f(t)$ is the {\it time-domain filter function} that encapsulate the effects of the periodic pulse sequence described above \cite{deSousa_TAP09,Cywinski_PRB08,Szankowski_JPCM17,Szankowski_PRA18,Krzywda_PRA17}. This filter function is depicted in Fig.~\ref{fig:f_harmonics}: it has a form of a square wave, that switches between $+1$ and $-1$ at the moment that coincides with the timing of pulse application. Due to {\it periodicity} of the sequence, the resultant filter function oscillates with a well-defined frequency
\begin{equation}
\omega_p = \frac{\pi}{\tau_p}.
\end{equation}
Therefore, $f(t)$ acts as a frequency filter that allows to pass through only the harmonics of the noise that are commensurate with the frequency of the filter's oscillations. The idea is that, by scanning the wide range of settings of pulse sequence periods and measuring the corresponding qubit response, it is possible to reconstruct the frequency distribution of the noise, i.e., to perform the {\it noise spectroscopy} \cite{Alvarez_PRL11,Bylander_NP11,Yuge_PRL11,Szankowski_JPCM17}.

The single-qubit method described above is, naturally, the {\it temporal} type of spectroscopy that can only acquire information concerning the temporal fluctuations of the noise. Here we propose a scheme for performing the full {\it spatiotemporal} spectroscopy that provide a more complete characterization of the noise field with a description of both its spatial and temporal fluctuations. The design of this type of spectrometer is derived from the results of Refs.~\cite{Szankowski_PRA16,Paz_NJP16,Krzywda_PRA17}, where it was shown that in order to gain access to non-trivial information of this nature one requires a probe composed of multiple spatially distributed qubits that are driven with properly coordinated pulse sequences, and it consists of three essential elements:
\begin{enumerate} 
\item The spectrometer is to be constructed out of multiple, non-interacting qubits, labeled with index $q=1,\ldots,N$, that are arranged in a line--which is the natural geometry for ion trap qubits \cite{Monz_PRL11}, and which is pursued in recent experiments with quantum dot spin qubits \cite{Zajac_PRAPL16,Fujita_NPJQI17,Malinowski_arXiv18}---i.e., qubits' positions satisfy
\begin{equation}
\mathbf{r}_q = x_q \mathbf{n},
\end{equation}
where $\mathbf{n}$ is the unit vector that establishes the spatial orientation of the spectrometer. The set of coordinates $x_q$ defines the qubit {\it spatial distribution}
\begin{equation}
\rho(x) = \sum_{q=1}^N \delta(x-x_q)\,,
\end{equation}
where $\delta(z)$ is the Dirac delta function. For such a geometry of the register the multi-qubit phase Hamiltonian is given by $\hat V_{N\mathrm{Q}} = \sum_{q=1}^N \xi(x_q\mathbf{n},t)\hat\sigma_z^{(q)}/2$, which commutes with the free Hamiltonian $\hat H_0 = \sum_{q=1}^N \Omega_q\hat\sigma_z^{(q)}/2$.

\item  The qubits can be addressed with individually tailored pulse sequences that produce the qubit-wise time-domain filter functions $f_q(t)$. We require that these filter functions satisfy
\begin{equation}
f_q(t) = f(t+\Delta\tau_q)\,,
\end{equation}
and that the time shift of qubit's filter is associated with its {\it physical position} via linear relation
\begin{equation}\label{eq:kp_def}
\Delta\tau_q = \frac{k_p}{\omega_p} x_q\,,
\end{equation}
and $k_p$ is a real, adjustable parameter, that plays the role analogous to $\omega_p$.

\item The linear geometry of qubit register coupled with time-shifted pulse sequences are sufficient to produce spatial and temporal frequency filters for the noise field. However, in order to accurately retrieve the resultant spectroscopic information it is crucial that the qubits' spatial distribution forms a pattern of periodically repeated blocks, although the distribution withing a single block does not have to be so ordered. Hence, we require that the total number of qubits is $N=n_\mathrm{s} N_0$, where the integer $n_\mathrm{s}\geqslant 1$, $N_0$ is the number of qubits constituting a single block, and the spatial distribution can be written as
\begin{align}
\label{eq:periodic_rho}
\rho(x) = \sum_{r=0}^{n_\mathrm{s}-1}\rho_0(x-r L_0)\,.
\end{align}
Here $L_0$ is the distribution period, and $\rho_0(x)=\sum_{q=1}^{N_0}\delta(x-x_q)$ is the spatial distribution of the first block of qubits. For convenience we choose the origin of the coordinate system so that $0<x_1<x_2<\ldots<x_{N_0}$ and $L_0=x_{N_0}+x_1$. The period of distribution can be identified with the length of the qubit block; consequently, the {\it total length} of the spectrometer is then given by 
\begin{equation}
L=n_\mathrm{s}L_0.
\end{equation}
The simplest example of qubit spatial distribution that adheres to \eqref{eq:periodic_rho} constitutes of $N$ equidistant qubits, which is equivalent to spectrometer composed of $N$ single qubit blocks. However, as we argue in Sec.~\ref{sec:examples}, blocks containing larger numbers of qubits with more disordered distributions allow for more precise spectroscopy in most cases.
\end{enumerate}

\begin{figure}[t]
\centering
\includegraphics[width=\columnwidth]{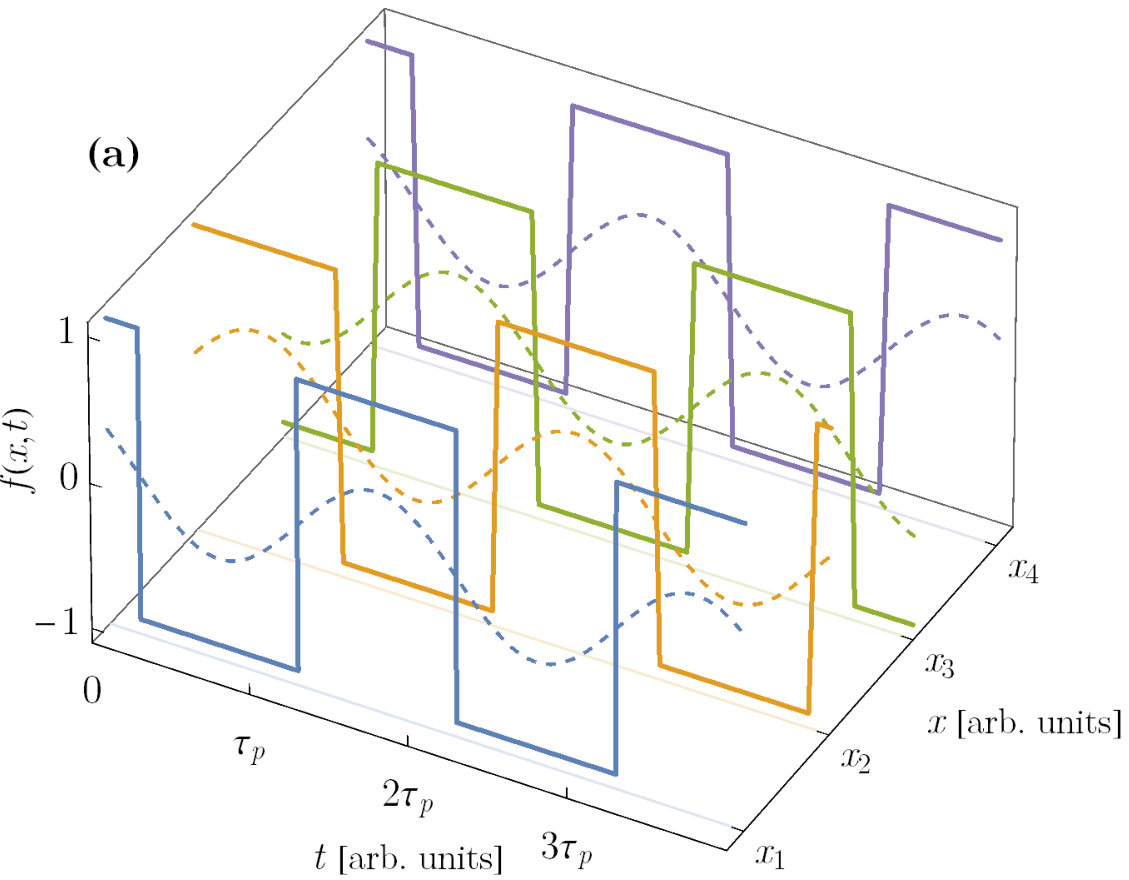}\vspace{.2cm}
\includegraphics[width=\columnwidth]{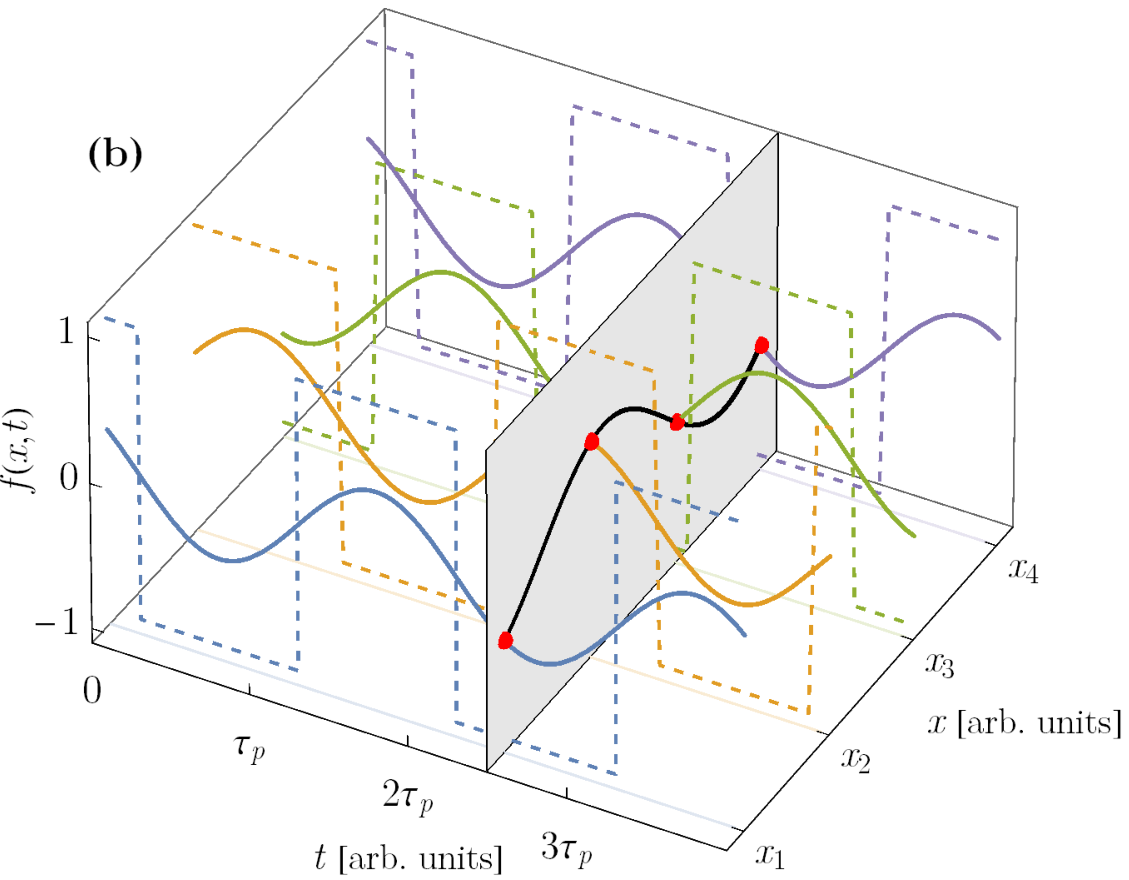}
\caption{(a) The spatiotemporal filter $f(x,t)$: it is composed of a number of time-domain filter functions $f(t)$ (one for each spectrometer qubit, hence the discretization of spatial degree of freedom) that are time-shifted by an amount proportional to the spatial argument $x=x_1,x_2,\ldots,x_N$, where $x_q$ is the position of $q$th qubit. The first harmonic of the constituent time-domain filter functions are also indicated (compare with Fig.~\ref{fig:f_harmonics}). (b) The cross section along the spatial direction of the first harmonic of each time-domain filter shows that the resultant points (red dots) form a pattern that can be fitted with $\propto\sin(k_p x+\varphi)$ (black line), where $k_p$ is the proportionality factor determining the time-shifts of the constituent filters [see \eqref{eq:kp_def}].}
\label{fig:st_comb}
\end{figure}

With the above requirements met, the evolution of multi-qubit ladder operator is given by
\begin{align}
\nonumber&(\hat\sigma_+^{(1)}\otimes\ldots\otimes\hat{\sigma}_+^{(N)})(T)=\\
\nonumber
&\\[-0.3cm]
\label{eq:N_coherence}
&=\overline{\exp\left[-i \sum_{q=1}^N \phi_\xi^{(q)}(T)\right]}\,\hat\sigma_+^{(1)}\otimes\ldots\otimes\hat\sigma_+^{(N)}\,,
\end{align}
with the phase acquired for a single noise realization given by
\begin{align}
\nonumber
&\sum_{q=1}^N \phi^{(q)}_\xi (T)= \int_0^T dt \sum_{q=1}^N f(t+\Delta\tau_q)\xi(x_q \mathbf{n},t)\\
\label{eq:N_phase_int_form}
&\equiv \int_0^L dx \int_0^{T} dt \, f(x,t)\xi(x\mathbf{n},t)\,,
\end{align}
where  the {\it spatiotemporal filter function}
\begin{equation}\label{eq:xt_filter}
f(x,t)=f\left(t+\frac{k_p}{\omega_p} x\right)\rho(x)
\end{equation}
is illustrated in Fig.~\ref{fig:st_comb} (a).

As it is highlighted in Fig.~\ref{fig:st_comb} (b), when one examines the cross section along the spatial dimension of a given temporal harmonic (compare with Fig.~\ref{fig:f_harmonics}) the sinusoidal waveform, with wavenumber $k_p$, is revealed. Therefore, it stands to reason that $f(x,t)$ would act as a frequency filter for both temporal and spatial directions; Sec.~\ref{sec:tech} is dedicated to proper formal proof that this is the case. This is the main qualitative results of this work.

However, due to the specifics of its physical implementation, the pulse-sequence-generated filters are burdened by a number of limitations. (i) Since a square wave itself can be decomposed into superposition of sine waves (see Fig.~\ref{fig:f_harmonics}), the temporal part of the frequency filter possesses infinitely (but countably) many passbands centered at the odd multiples of the filter frequency $\omega_p$. (ii) Although the spatial waveform is sinusoidal, one cannot avoid the appearance of side passbands, because of the discrete nature of filter function in $x$ direction. (iii) Due to finite duration of the evolution and the length of the spectrometer, both temporal and spatial passbands have finite width, proportional to $T^{-1}$ in the former case, and to $L^{-1}$ in the latter. 

In order to overcome these imperfections of this type of filter, so that an accurate spatiotemporal spectroscopy can be carried out, one has to implement a generalized version of the procedure for data acquisition and analysis, that was originally developed for single-qubit spectrometers in \cite{Szankowski_PRA18}. The details of this procedure are discussed in Sec.~\ref{sec:dist_proc} where we show that the periodicity of qubit spatial distribution and of quibt-wise pulse sequences are indispensable for its successful execution.

The method of spatiotemporal spectroscopy presented in this work is most suited for characterizing {\it stationary} and spatially {\it uniform} Gaussian noise fields (or when the Gaussian approximation is adequate to describe qubit-noise coupling \cite{Szankowski_JPCM17}). In such a case, all statistical properties of $\xi$ are defined by the average value and the auto-correlation function (Gaussianity), that is respectively constant and depends only on the relative position and time (stationarity and uniformity),
\begin{align}
&\overline{\xi(\mathbf{r},t)} = \xi_0\,,\\[.2cm]
&\overline{\xi(\mathbf{r},t)\xi(\mathbf{r}',t')}-\xi_0^2 = C(\mathbf{r}-\mathbf{r}',t-t')\,.
\end{align}
Then, the final product of the method is reconstruction of the spectral density, that describes the average distribution of power among spatiotemporal harmonic components of the noise,
\begin{equation}\label{eq:spec}
S_\mathbf{n}(k,\omega) = \int_{-\infty}^\infty dx\, dt\,C(x\mathbf{n},t)e^{-i k x-i \omega t}\,.
\end{equation}
The subscript $\mathbf{n}$ indicates that such a spectrum quantifies fluctuation only in one spatial dimension determined by the physical orientation of the spectrometer.

\section{Coordinated pulse sequences as a spatiotemporal frequency filter}\label{sec:tech}
The goal of this sections is to demonstrate formally that the spatiotemporal filter function $f(x,t)$, introduced in \eqref{eq:xt_filter}, acts as a two-dimensional passband frequency and wavenumber filter of the spectral density $S_\mathbf{n}(k,\omega)$.
\begin{figure}[t]
\centering
\includegraphics[width=\columnwidth]{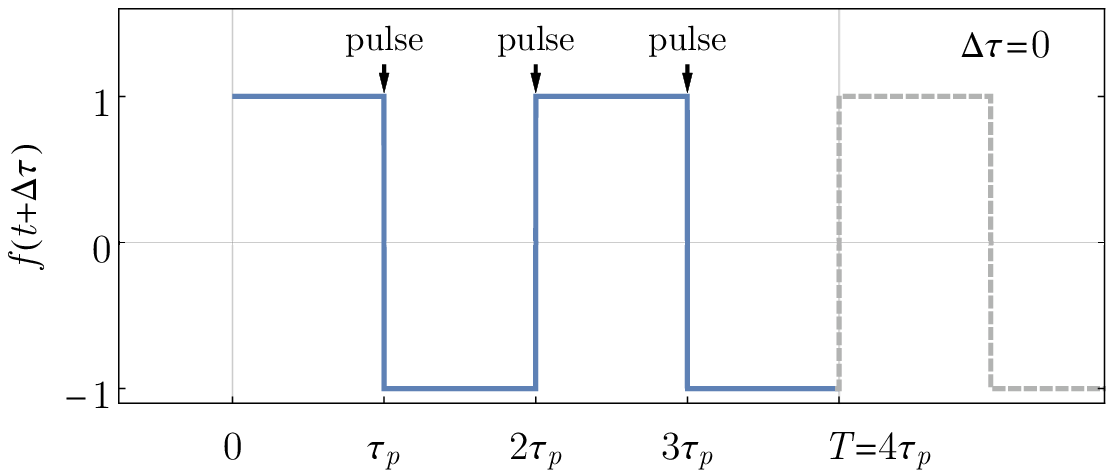}\vspace{.1cm}
\includegraphics[width=\columnwidth]{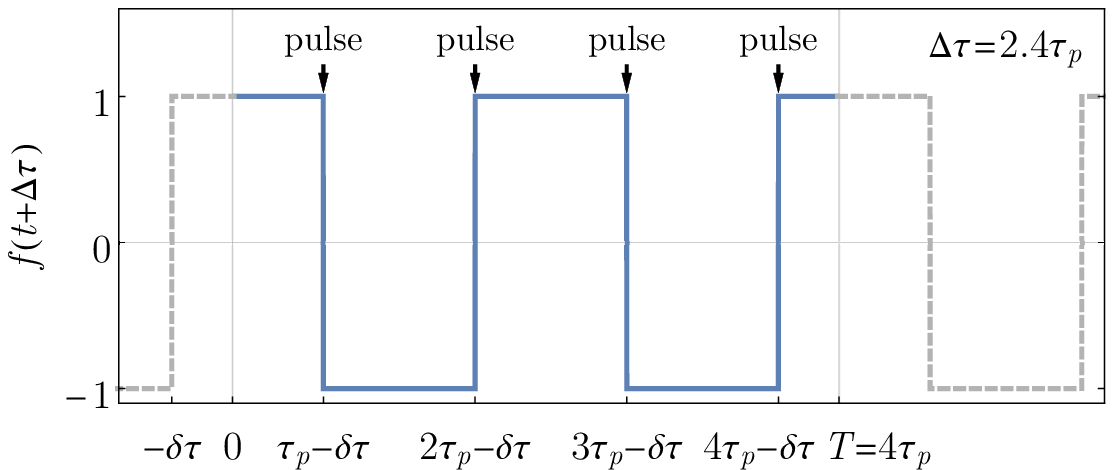}\vspace{.1cm}
\includegraphics[width=\columnwidth]{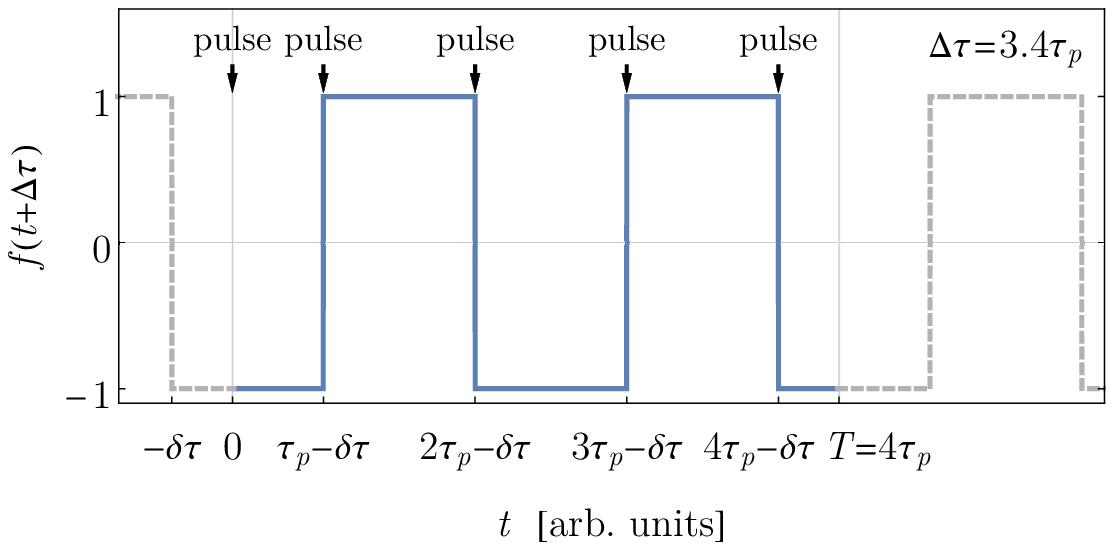}
\caption{The time-shifted filter functions $f(t+\Delta\tau)$, for three instances of the value of the time-shift $\Delta\tau$: (from top to bottom) $0$, $2.4\tau_p=2\tau_p+\delta\tau$ and $3.4\tau_p=3\tau_p+\delta\tau$. The blue solid line highlights the segment of filter functions within the interval of the evolution, $t\in[0,T=4\tau_p]$, while the gray dashed line indicates the portions of filters that lie beyond this interval. The timings of pulses in sequences producing the filter functions coincide with their sign switches, with an exception of the bottommost case, where an additional pulse is applied at the beginning of the evolution so that the filter at $t=0$ starts at $-1$ value.}
\label{fig:shifted_seq}
\end{figure}

\subsection{The attenuation function}
The figure of merit of noise spectroscopy is so-called {\it attenuation function} $\chi(L,T)$, defined as
\begin{equation}\label{eq:att_fnc_def}
e^{-\chi(L,T)}\equiv \overline{\exp\left(-i\int_0^L \!\!dx\int_0^T\!\!dtf(x,t)\xi(x\mathbf{n},t)\right)} \,\, ,
\end{equation}
where the relation \eqref{eq:N_phase_int_form} has been invoked to express the stochastic phase in terms of spatiotemporal filter function. The attenuation function is considered as a raw data point obtained from direct measurement of the multi-qubit spectrometer response to noise field [see \eqref{eq:N_coherence}], conditioned by the settings of applied pulse sequences. The processing of this data in order to reconstruct the spectral density is what constitutes the core of the method.

For noise field with Gaussian statistics the average in \eqref{eq:att_fnc_def} can be carried out to yield a closed form expression
\begin{align}
\nonumber
&\overline{\exp\left(-i\int_0^L\!\!dx\int_0^T\!\!dt\,f(x,t)\xi(x\mathbf{n},t)\right)} \\
\nonumber &\\[-.4cm]
&= \exp\left[ -\frac{1}{2}\overline{\Bigg(\int_0^L \!\!dx\int_0^T\!\!dt\, f(x,t)\xi(x\mathbf{n},t)\Bigg)^2}\, \right].
\end{align}
Utilizing the definition of auto-correlation function, and its relation to spectral density, the attenuation function can be transformed into the following form
\begin{align}
\label{eq:att_func_ft}
\chi(L,T) = \frac{1}{2}\int_{-\infty}^\infty \frac{d k}{2\pi}\frac{d\omega}{2\pi} | \tilde{f}_{L;T}(k,\omega)|^2 S_\mathbf{n}(k,\omega)\,,
\end{align}
where
\begin{align}
\nonumber
\tilde{f}_{L;T}(k,\omega) &= \int_{-\infty}^\infty dxdt\,e^{-ikx-i\omega t}\\
\label{eq:spf_ft_def}
&\times\Theta(L-x)\Theta(x)\Theta(T-t)\Theta(t)f(x,t)\,,
\end{align}
is the Fourier transform of spatiotemporal filter function restricted to the duration of the evolution and the length of the spectrometer.

\subsection{Pulse sequence as a frequency filter}\label{sec:DD_as_filter}
Here we formally define the spatiotemporal filter function $f(x,t)$, and provide the result for its two-dimensional Fourier transform \eqref{eq:spf_ft_def}.

Since the time-domain filter function $f(t)$, produced by the periodic sequence composed of odd $2n_\mathrm{t}-1$ number of pulses, applied over the duration that encompasses all these periods, 
\begin{equation}
T=n_\mathrm{t}2\tau_p=n_\mathrm{t}T_0,
\end{equation}
does not appear anywhere outside the integral in \eqref{eq:1_qubit_phase}, it is formally not required to specify it outside the interval $[0,T]$; however it is convenient to do it anyway. Here we choose the definition where the filter function is extended beyond the duration of the evolution by repeating its period indefinitely, 
\begin{equation}\label{eq:f_PDD}
f(t)=\sum_{j=0}^{\infty}(-1)^j\Theta((j+1)\tau_p-t)\Theta(t-j\tau_p),
\end{equation}
where $\Theta(t)$ is the Heaviside step function that is $1$ for $t>0$ and $0$ otherwise.

With definition \eqref{eq:f_PDD} in hand, the qubit-wise filter functions are straightforward to model mathematically as $f_q(t) = f(t+\Delta\tau_q)$ (with $\Delta\tau_q\geq0$), however the design of pulse sequences that would produce them, require some explanation, see Fig.~\ref{fig:shifted_seq}. The filter function with zero-shift, $f(t)$, is the result of a sequence where a pulse is applied every $\tau_p$, for the total of $2n_\mathrm{t}-1$ pulses over the duration $T=n_\mathrm{t} T_0$. For the same duration and the shift that satisfies $\Delta\tau_q =2\kappa \tau_p + \delta\tau$, with $\kappa\in\text{Integers}$ and $0<\delta\tau\leqslant\tau_p$, the required sequence consists of the first pulse timed at $\tau_p-\delta\tau$, followed by $2n_\mathrm{t}-1$ pulses applied every $\tau_p$. Finally, when $\Delta\tau_q =(2\kappa+1)\tau_p + \delta\tau$, the pattern of pulse timings is the same as in the previous case, with an addition of one more pulse applied at $t=0$, so that the shifted filter function begins with the negative value.

Now we proceed to find the Fourier transform of restricted spatiotemporal filter \eqref{eq:spf_ft_def} (the details of the calculations can be found in Appendix \ref{app:spf_ff}):
\begin{align}
\nonumber
\tilde{f}_{L;T}(k,\omega)&=\tilde{f}_{n_\mathrm{s}L_0; n_\mathrm{t}T_0}(k,\omega)\\
\nonumber
&=\sum_{m=-\infty}^\infty c_{m\omega_p} h_{n_\mathrm{t}T_0}(\omega-m\omega_p)\\
\label{eq:spf_ft}
&\phantom{=}\times\sum_{l=-\infty}^\infty \csp{lk_d} h_{n_\mathrm{s}L_0}(k-mk_p+lk_d),
\end{align}
where $\omega_p=\pi/\tau_p$ is the filter frequency and $k_d = 2\pi/L_0$ is the wavenumber associated with the period of qubit spatial distribution. The shape of the passband of each harmonic component of the filter is given by
\begin{align}
\nonumber
h_W(u) &= \int_{-\infty}^\infty dw\,e^{-i w u}\Theta(W-w)\Theta(w)\\
\label{eq: sincs}
&= \frac{W}{2}e^{i \frac{W u}{2}}\sinc\left(\frac{W u}{2}\right)\,,
\end{align}
and the corresponding weights equal to
\begin{align}
\label{eq:c}
c_{m\omega_p} &=\frac{1}{T_0}\int_0^{T_0}\!\!\!\!e^{-i m \omega_p t}f(t)dt
	=\left\{\begin{array}{cl} \frac{2}{i m \pi} &\text{for $m\in\,$Odd}\\[.2cm] 0 &\text{otherwise}\\\end{array}\right.,\\[.3cm]
\label{eq:s}
\csp{l k_d} &=\frac{1}{L_0}\int_{0}^{L_0} \!\!\!\!e^{-i  l k_d x}\rho_0(x)dx = \frac{k_d}{2\pi}\sum_{q=1}^{N_0} e^{-i l k_d x_q}.
\end{align}
They can be identified with the Fourier series coefficients of the time-domain filter function \cite{Szankowski_JPCM17,Krzywda_PRA17,Szankowski_PRA18}, and of the spatial distribution of qubit block. Note how only the widths of passbands depend on the number of repetitions of qubit blocks $n_\mathrm{s}$ or the number of pulses $n_\mathrm{t}$; this is a direct consequence of periodicity of qubit spatial distribution and pulse sequences used for construction of the spectrometer.

The \eqref{eq:spf_ft} readily confirms the wavenumber and frequency filtering properties of $f(x,t)$ described in Sec.~\ref{sec:intro}: (i) The frequency-$\omega$- and wavenumber-$k$-dependent $h$ functions define the temporal and spatial passbands of the filter. Their respective widths are $T^{-1}$ and $L^{-1}$. (ii) Due to harmonic distribution of $f(t)$---quantified by Fourier coefficients $c_{m\omega_p}$---the temporal passbands are located around odd multiples of $\omega_p$. (iii) The temporal part of the filter interferes with the spatial part by shifting the positions of its passbands, so that for each temporal harmonic $c_{m\omega_p}$ there is a whole set of spatial passbands located around wavenumbers $mk_p-l k_d$, each one weighted by the Fourier series coefficient $\csp{l k_d}$.

\section{Spatiotemporal noise spectroscopy}\label{sec:dist_proc}

Below we present the detailed overview of the procedure of spectral density reconstruction with an accompanying discussion of its implementation.

\subsection{The spectroscopic formulas}\label{sec:spec_formula}

It has been demonstrated in Ref.~\cite{Szankowski_PRA18} that the overlap integral of the form 
\begin{equation}\label{eq:att_fnc_form}
\frac{1}{2}\int_{-\infty}^\infty \frac{d\omega}{2\pi}\big|\sum_{m} g_{m} h_{n_\mathrm{t}T_0}(\omega-\omega_m)\big|^2 F(\omega)\,,
\end{equation}
with $F(\omega)$ being continuous and non-negative function (like any spectral density), can be broken down into combination of terms that are subject to three distinct types of scaling laws in respect to $n_\mathrm{t}$. If one substitutes $g_m = c_{m\omega_p}\sum_{l}\csp{lk_d}h_{n_\mathrm{s}L_0}(k-mk_p+lk_d)$, $\omega_m=m\omega_p$, and $F(\omega)=S_{\mathbf{n}}(k,\omega)$ in \eqref{eq:att_fnc_form}, this theorem also applies to the case of attenuation function $\chi(L,T)=\chi(n_\mathrm{s}L_0,n_\mathrm{t}T_0)$, for which the following decomposition is obtained:
\begin{equation}\label{eq:T-scale_laws}
\chi(n_\mathrm{s}L_0,n_\mathrm{t}T_0) = \chi_\mathrm{S}+\Delta\chi_T+\Delta\chi_0\,,
\end{equation}
where $\chi_\mathrm{S}\propto n_\mathrm{t}$, $\Delta\chi_0$ is independent of $n_\mathrm{t}$, and $\Delta\chi_T\sim C(0,n_\mathrm{t}T_0)$.

The term that scales linearly with $n_\mathrm{t}$, the so-called {\it spectroscopic formula} $\chi_\mathrm{S}$, is given by
\begin{align}
\nonumber
&\chi_\mathrm{S} = \frac{n_\mathrm{t}T_0}{2}\sum_{m=-\infty}^\infty |c_{m\omega_p}|^2 \\
\nonumber
&\times\int_{-\infty}^\infty 
	\Big|\sum_{l=-\infty}^\infty \csp{l k_d}h_{n_\mathrm{s}L_0}(k-mk_p+lk_d)\Big|^2
	\, S_\mathbf{n}(k,m\omega_p)\frac{dk}{2\pi}\\
\label{eq:spec_formula}
&\phantom{\chi_\mathrm{S}}\equiv \frac{n_\mathrm{t}T_0}{2}\sum_{m={-}\infty}^\infty |c_{m\omega_p}|^2 S^\star(m\omega_p|n_\mathrm{s}L_0).
\end{align}
It has a form of marginal spectral density $S^\star(m\omega|L)$ (i.e.,~spectral density integrated over spatial degree of freedom) spanned on a frequency comb, the teeth of which are positioned at the central frequencies of temporal passbands. In the case of single-qubit noise spectroscopy where only the temporal degree of freedom is analyzed, so that $S^\star$ is replaced with purely temporal spectral density, $\chi_\mathrm{S}$ is the main resource for spectrum reconstruction; it establishes a simple relation between the measured value of attenuation function for a given setting of $\omega_p$ and the values of spectral density. By combining the information from a set of spectroscopic formulas obtained with properly chosen settings of filter frequencies this relation can be inverted, and thus the spectrum of temporal fluctuations is reconstructed \cite{Alvarez_PRL11,Szankowski_PRA18}. In the case of multi-qubit spectrometer it is only an intermediate step, as $S^\star$ can be interrogated further to extract information on spatiotemporal spectral density $S_\mathbf{n}(k,\omega)$.

Terms $\Delta\chi_T$ and $\Delta\chi_0$ are corrections to spectroscopic formula due to finite widths of the temporal passbands \cite{Szankowski_PRA18}. They diminish accuracy of the spectroscopy, by deviating the attenuation function from the expected form of spectrum-spanned-on-frequency comb required by the reconstruction procedure to recreate spectral density with satisfactory level of fidelity. The $n_\mathrm{t}$-independent $\Delta\chi_0$ can be effectively eliminated using the method described in the upcoming Sec.~\ref{sec:lin_fit}. On the other hand, the remaining $\Delta\chi_T\sim C(0,n_\mathrm{t}T_0)$ cannot be treated in this way. Instead, it can be made negligible (in comparison to e.g., the measurement error) by exploiting the properties of the auto-correlation function that determine the $n_\mathrm{t}$-scaling law of $\Delta\chi_T$. It is expected on physical grounds that the noise field auto-correlation function has a finite range, both in temporal and spatial dimension, i.e., there exist such a scale of time $t_c$ and length $x_c$ (called the {\it correlation time} and {\it correlation length}, respectively) that
\begin{equation}
\label{eq:corr_time_and_length}
\begin{array}{ll}
C(x\mathbf{n},t)\xrightarrow{|t|\gg t_c} 0&\ \ \text{for any $x$, or }\\
C(x\mathbf{n},t)\xrightarrow{|x|\gg x_c} 0&\ \ \text{for any $t$.}\\
\end{array}
\end{equation}
Therefore, setting the number of pulses $n_\mathrm{t}$ so that the duration $T=n_\mathrm{t}T_0$ is much longer than the correlation time and $C(0,n_\mathrm{t} T_0)\to 0$, is sufficient to meet the requirement that the $n_\mathrm{t}$-dependent correction term is negligible.

Since the form of $S^\star$ matches that of \eqref{eq:att_fnc_form}, the reasoning presented in Ref.~\cite{Szankowski_PRA18} that lead to $n_\mathrm{t}$-scaling laws of different parts of attenuation function, can be directly applied to the marginal spectral density as well. Therefore, $S^\star$ also decomposes into three terms that correspond to \eqref{eq:T-scale_laws}, but with the spectrometer length $L=n_\mathrm{s}L_0$ replacing the duration $T=n_\mathrm{t}T_0$:
\begin{equation}\label{eq:S_star_L-scaling_decomposition}
S^\star(m\omega_p|n_\mathrm{s}L_0)=S^\star_\mathrm{S}+\Delta S^\star_0+\Delta S^\star_L,
\end{equation}
where the $n_\mathrm{s}$-independent $\Delta S_0^\star$ and $\Delta S^\star_L\sim\int_{-\infty}^\infty  e^{-i m\omega_p t}C(n_\mathrm{s}L_0,t) dt$, are the corrections to the spatial version of spectroscopic formula,
\begin{equation}
\label{eq:spat_spec_formula}
S^\star_\mathrm{S}=n_\mathrm{s}L_0\sum_{l=-\infty}^\infty|\csp{l k_d}|^2 S_\mathbf{n}(mk_p-lk_d,m\omega_p),
\end{equation}
that enables the full reconstruction of spectral density $S_\mathbf{n}(k,\omega)$. In a manner analogous to \eqref{eq:T-scale_laws}, meeting the condition $n_\mathrm{s}L_0\gg x_c$ is sufficient to treat the $n_\mathrm{s}$-dependent correction as negligible, while the constant $\Delta S_0^\star$ can be taken care of with the method of Sec.~\ref{sec:lin_fit}, which we now proceed to discuss.

\subsection{The linear fit method for acquiring the spectroscopic formulas}\label{sec:lin_fit}

In order to perform an accurate noise spectroscopy it is crucial to gain access to parts of the attenuation function $\chi(n_\mathrm{s}L_0,n_\mathrm{t}T_0)$ that are described by spectroscopic formulas $\chi_\mathrm{S}$ and $S^\star_\mathrm{S}$ [see Eqs.~\eqref{eq:spec_formula} and \eqref{eq:spat_spec_formula}]. The description of \'{A}lvarez-Suter method \cite{Alvarez_PRL11} for extracting the values of spectral density from $\chi_\mathrm{S}$ and $S^\star_\mathrm{S}$ by deconvolving them from the frequency combs they are spanned on is postponed to Sec.~\ref{sec:AS_method}. Here we discuss an application of a scheme that allows to extract these key parts of $\chi(n_\mathrm{s}L_0,n_\mathrm{t}T_0)$ by exploiting the differences in the scaling laws of its constituents.

The first step is to generate a dataset composed of attenuation functions gathered in a series of measurements where the number of qubit block repetitions $n_\mathrm{s}$ and pulses $n_\mathrm{t}$ are varied in each experiment, starting from the lowest $n_\mathrm{s}^{\mathrm{min}}$ and $n_\mathrm{t}^\mathrm{min}$, and progressively increased to $n_\mathrm{s}^\mathrm{max}$ and $n_\mathrm{t}^\mathrm{max}$. On one hand, the required control over the number of applied pulses is both conceptually and practically trivial. On the other hand, an analogous method for extending the length of the spectrometer that involves adding blocks of qubits into the system would be substantially more challenging to implement than any manipulation of the pulse sequences. Therefore, we propose a different, but effectively equivalent approach where the number of qubits is constant, and instead of physically removing or adding qubits to the system one can turn on and off the contribution from specific blocks by focusing on suitable observables. In particular, in the Heisenberg picture the operator
\begin{align}
\nonumber
(\hat\sigma_+^{(1)}\otimes\ldots\hat\sigma_+^{(r N_0)}\otimes\hat{\mathds{1}}^{(r N_0+1)}\otimes\ldots\hat{\mathds{1}}^{(N)})(T) =&\\
\nonumber
=\hat\sigma_+^{(1)}\otimes\ldots\hat\sigma_+^{(r N_0)}\otimes\hat{\mathds{1}}^{(r N_0+1)}\otimes\ldots\hat{\mathds{1}}^{(N)}&\\
\label{eq:H_pic}
\phantom{=}\times\overline{\exp\left(-i \sum_{q=1}^{r N_0}\phi^{(q)}_\xi(T)\right)}&
\end{align}
evolves only due to contribution from the qubits belonging to the first $r$ blocks. The attenuation function encoded in operator of type \eqref{eq:H_pic} is accessed via corresponding set of tomographic measurements performed on spectrometer qubits that were initialized in an adequately prepared state. For such an interrogation to be successful it is necessary that the state, described by the multi-qubit density matrix $\hat\varrho$, satisfies
\begin{align}
\nonumber
&\mathrm{Tr}_{1\ldots N}\left[\,\hat\varrho\,\hat\sigma_+^{(1)}\otimes\ldots\hat\sigma_+^{(rN_0)}\otimes\hat{\mathds{1}}^{(rN_0+1)}\otimes\ldots\hat{\mathds{1}}^{(N)}\right]
	\\[.2cm]
\nonumber
&=\mathrm{Tr}_{1\ldots rN_0}\left[\,(\mathrm{Tr}_{rN_0+1\ldots N}\hat\varrho)\hat\sigma_+^{(1)}\otimes\ldots\hat\sigma_+^{(rN_0)}\right]\\[.2cm]
&=\langle {-}z^{(1)}\ldots{-}z^{(rN_0)}|(\mathrm{Tr}_{rN_0+1\ldots N}\hat\varrho)|{z}^{(1)}\ldots{z}^{(rN_0)}\rangle \neq 0\,,
\end{align}
where $\mathrm{Tr}_{q_1q_2\ldots}$ is a trace over degrees of freedom of qubits $q_1,q_2,\ldots$, and $|\pm z^{(q)}\rangle$ are the eigenstates of $\hat\sigma^{(q)}_z$. This requirement is met by, e.g., qubits initialized in a separable state $\hat\varrho=|x^{(1)}\rangle\langle x^{(1)}|\otimes\ldots |x^{(N)}\rangle\langle x^{(N)}|$, $\hat\sigma^{(q)}_x|\pm x^{(q)}\rangle=\pm|\pm x^{(q)}\rangle$, for which the above matrix element equals $2^{-rN_0}$. Much larger value of $2^{-1}$ is obtained for an entangled state of a form $\hat\varrho=|\Psi_{\mathrm{GHZ}}^{(1\ldots rN_0)}\rangle\langle\Psi_{\mathrm{GHZ}}^{(1\ldots rN_0)}|\otimes\hat\varrho^{(rN_0+1\ldots N)}$, where
\begin{equation}\label{eq:GHZ}
|\Psi_{\mathrm{GHZ}}^{(1\ldots rN_0)}\rangle = \frac{|{-z^{(1)}}\ldots{-z^{(rN_0)}}\rangle+|z^{(1)}\ldots z^{(rN_0)}\rangle}{\sqrt 2}\,,
\end{equation}
and $\hat\varrho^{(rN_0+1\ldots N)}$ is an arbitrary density matrix of qubits $rN_0+1,\ldots,N$. These examples show that qubits composing the spectrometer does not necessary have to be correlated, but the correlations---and in particular, the entanglement of the type present in states \eqref{eq:GHZ}---can be beneficial, as they increase the magnitude of the measured signal and thus enhance the relative accuracy of tomographic measurements required for extracting the raw spectroscopic data. Such states of multiple qubits were created in ion traps \cite{Leibfried_Nature05,Monz_PRL11} and with superconducting qubits \cite{Barends_Nature14,Song_PRL17}.

The gathered attenuation functions can be conveniently arranged into $(n_\mathrm{s}^\mathrm{max}-n_\mathrm{s}^\mathrm{min}+1)\times (n_\mathrm{t}^\mathrm{max}-n_\mathrm{t}^\mathrm{min}+1)$ matrix:
\begin{align}
\label{eq:data}
\nonumber
&[\chi_{n_\mathrm{s},n_\mathrm{t}}] =\\
\nonumber
&= [\chi\big(n_\mathrm{s}L_{0},n_\mathrm{t}T_{0})]_{(
	n_\mathrm{s}=n_\mathrm{s}^\mathrm{min},\ldots,n_\mathrm{s}^\mathrm{max}
	;\,n_\mathrm{t}=n_\mathrm{t}^\mathrm{min},\ldots,n_\mathrm{t}^\mathrm{max})}\\
&=\left[\begin{array}{ccc}
&&\\[-.25cm]
\chi(n_\mathrm{s}^\mathrm{min}L_0,n_\mathrm{t}^\mathrm{min}T_0) & \ldots & \chi(n_\mathrm{s}^\mathrm{min}L_0,n_\mathrm{t}^\mathrm{max}T_0) \\[.2cm]
\vdots            &          & \vdots \\[.2cm]
\chi(n_\mathrm{s}^\mathrm{max}L_{0},n_\mathrm{t}^\mathrm{min}T_0) & \ldots & \chi(n_\mathrm{s}^\mathrm{max}L_{0},n_\mathrm{t}^\mathrm{max}T_{0}) \\[.15cm]
\end{array}\right].
\end{align}

\begin{figure}[t]
\centering
\includegraphics[width=\columnwidth]{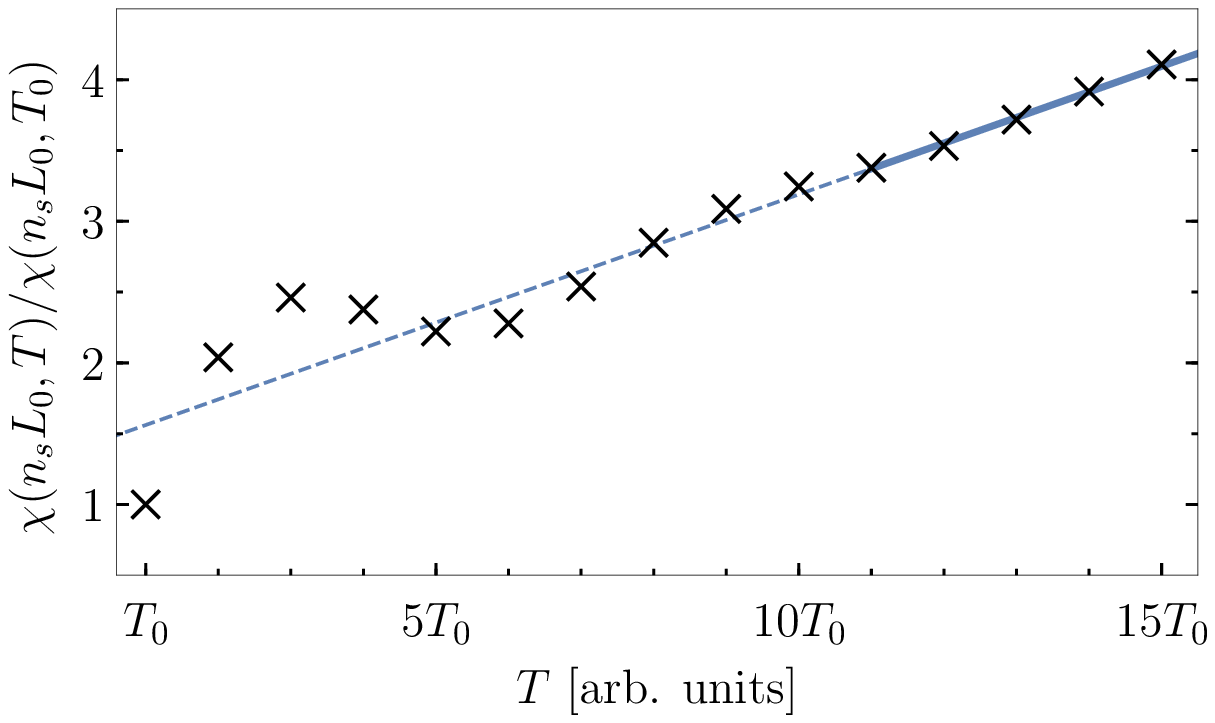}
\includegraphics[width=\columnwidth]{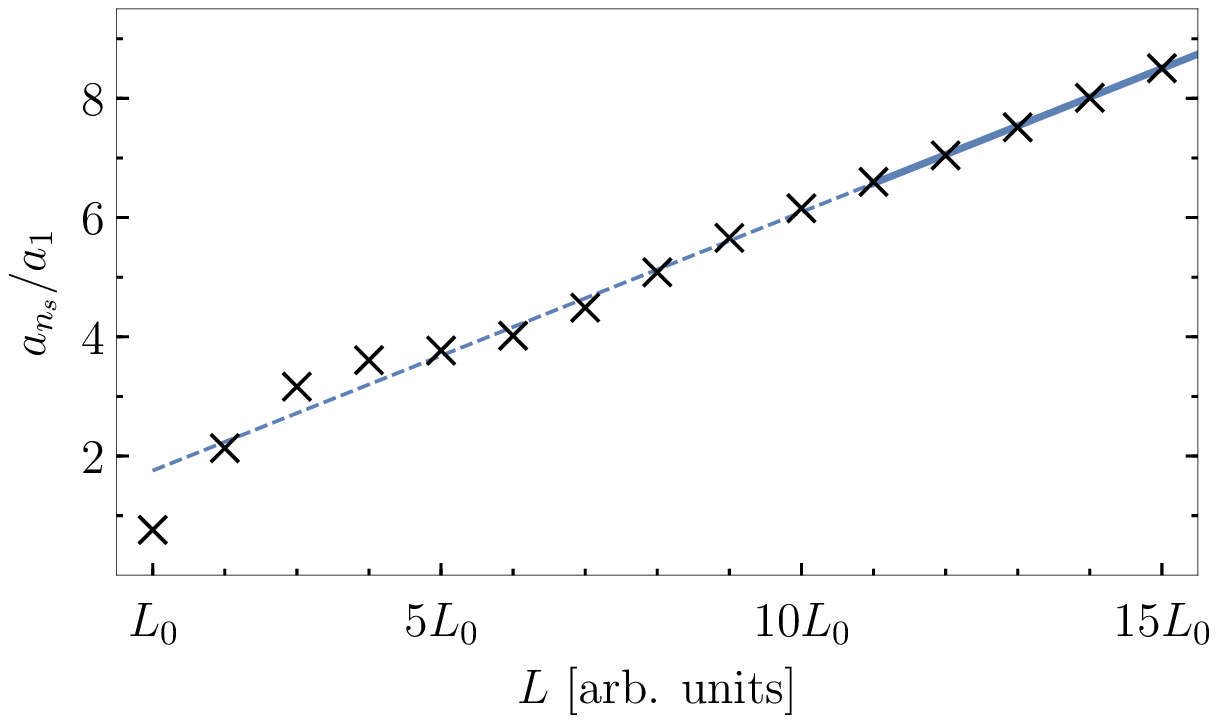}
\caption{Top figure: An example of graph composed of data set of attenuation functions $\{ \chi(n_\mathrm{s}L_0,T_0),\chi(n_\mathrm{s}L_0,2T_0),\ldots,\chi(n_\mathrm{s}L_0,15T_0)\}$ (here $n_\mathrm{s}=5$) plotted against the duration $T$ (black crosses). It features the initial oscillations [$T$ comparable with the correlation time $t_c$ so that $\Delta\chi_T\sim C(0,n_\mathrm{t} T_0)$ is relatively large, see \eqref{eq:corr_time_and_length}], that eventually decay away and transition into linear trend. The function $F(T)=a_{n_\mathrm{s}} T+b_{n_\mathrm{s}}$ is fitted to this emergent trend (blue, solid line). The cut off point for the data used for the fit is chosen so that it results in the best, in the least-squares sense, estimate of the trend's slope $a_{n_\mathrm{s}}$ and intercept $b_{n_\mathrm{s}}$. Bottom figure: The slopes obtained for different settings of $n_\mathrm{s}$, $\{ a_{1},\ldots,a_{15}\}$, are then plotted against the spectrometer length $L$. The course of the plot is similar to the previous case: the decaying oscillations transition into linear trend when $L\gg x_c$. The slope of this trend---which is determined by fitting $F(L)=AL+B$---is given by \eqref{eq:slope}. The measured attenuation functions were simulated by the exact numerical integration in \eqref{eq:att_func_ft}, with the spectral density $S_\mathbf{n}(k,\omega)$ taken to be of the same form as in the upcoming Sec.~\ref{sec:examples}; the filter frequency and wavenumber were set to $\omega_p=\pi/t_c$ and $k_p=\pi/x_c$, where $t_c$ and $x_c$ are the correlation time and length of the chosen spectrum. }
\label{fig:data_plot_chi_vs_T}
\end{figure}

In the second step of the procedure, for each $n_\mathrm{s}$ (i.e., for each row of the matrix), the data points $\{(n_\mathrm{t}^\mathrm{min}T_0, \chi_{n_\mathrm{s},n_\mathrm{t}^\mathrm{min}}),\ldots,(n_\mathrm{t}^\mathrm{max}T_{0},\chi_{n_\mathrm{s}, n_\mathrm{t}^\mathrm{max}})\}$ are plotted on the graph. An example of such a plot is presented in Fig.~\ref{fig:data_plot_chi_vs_T}. The initial portion of data points exhibit oscillatory behavior with visible decaying envelope---this indicates that the $n_\mathrm{t}$-dependent correction term $\Delta\chi_T\sim C(0,n_t T_0)$ is still active in this region; the time scale of its decay is the correlation time. Interestingly, the observation of finite frequency of oscillations of this term (which reflect the oscillations present in the auto-correlation function) indicates a non-zero characteristic frequency in the spectrum, i.e.,~the presence of maximum of spectral density (the spectral line) at finite frequency.
Note that it would be unreasonable to expect to be able to guess how long $n_\mathrm{t}^\mathrm{min}T_0$ should be to completely avoid the appearance of  $\Delta\chi_T$, since it is not guaranteed that one possesses prior knowledge about spectral density of the noise field, which include the correlation time $t_c$. In fact, it is the analysis performed here that allows to estimate $t_c$ (by observing the time scale of the envelope decay), and even the positions of the spectral lines (by analyzing the frequencies present in the oscillatory behavior), prior to complete reconstruction of the spectral density. Proceeding further along the plot, the duration becomes much longer than $t_c$, $\Delta\chi_T$ decays to nearly zero, and data points start to follow a clear linear trend.

The third step is to fit linear function $F(T)=a_{n_\mathrm{s}} T + b_{n_\mathrm{s}}$ to the part of the graph that is free of $n_\mathrm{t}$-dependent correction. The intercept of the fit is the $n_\mathrm{t}$-independent correction and the slope is the essential part of the spectroscopic formula
\begin{align}
a_{n_\mathrm{s}} = \frac{\chi_\mathrm{S}(T)}{T}= \frac{1}{2}\sum_{m=-\infty}^\infty |c_{m\omega_p}|^2 S^\star(m\omega_p|n_\mathrm{s}L_0) .
\end{align}
The intercepts $b_{n_\mathrm{s}}$ can be safely discarded, while the slopes $a_{n_\mathrm{s}}$ found for each $L=n_\mathrm{s}L_0$ are needed for further processing.

The fourth step is analogous to the second step, but instead of attenuation function vs. the duration, the graph is created out of a data set composed of the slopes obtained previously, $\{ (n_\mathrm{s}^\mathrm{min}L_0,a_{n_\mathrm{s}^\mathrm{min}}),\ldots,(n_\mathrm{s}^\mathrm{max}L_{0},a_{n_\mathrm{s}^\mathrm{max}})\}$. Since the behavior of $S^\star$ as a function of spectrometer length $n_\mathrm{s}L_0$ have the same features as that of $\chi$ as a function of duration $n_\mathrm{t}T_0$, the data points on their respective graphs would have a similar course, as it can be seen in Fig.~\ref{fig:data_plot_chi_vs_T}. Hence, the decay and possible oscillations of $n_\mathrm{s}$-dependent correction should be observed, and the linear trend  should emerge as $n_\mathrm{s}L_0$ becomes much longer than the correlation length $x_c$. The intercept of $F(L)=A L + B$ fitted to this trend is given by the $n_\mathrm{s}$-independent corrections, and it can be immediately discarded, while the slope contains the sought after spectroscopic formulas. Let us write it in a form that is fit as an input for spectral density reconstruction procedure:
\begin{align}
\nonumber
A(&k_p,\omega_p)=\frac{1}{2}\sum_{m=-\infty}^\infty |c_{m\omega_p}|^2 \frac{S_\mathrm{S}^\star(L)}{L}\\
\nonumber
&=\frac{1}{2}\sum_{m=-\infty}^\infty |c_{m\omega_p}|^2 \sum_{l=-\infty}^\infty
	|\csp{lk_d}|^2 S_\mathbf{n}(mk_p-lk_d,m\omega_p)\\
\label{eq:slope}
&=\sum_{m>0} |c_{m\omega_p}|^2\sum_{l=-\infty}^\infty |\csp{l k_d}|^2 S_\mathbf{n}(mk_p-lk_d,m\omega_p). 
\end{align}
Here, the symmetries of spectral density $S_\mathbf{n}(-k,-\omega) = S_\mathbf{n}(k,\omega)$, and Fourier coefficients $c_{-m\omega_p}=c^*_{m\omega_p}$, $\csp{-l k_d}=\csp{lk_d}^*$ have been used to limit the sum over $m$ to positive indices only. The arguments of  $A(k_p,\omega_p)$ have been shown explicitly to underline that the slope was generated for a given choice of spatiotemporal filter settings.

The method of linear fit presented here, aside from its main purpose of producing the spectroscopic formulas, also serves as a self-diagnostic tool for the spatiotemporal spectroscopy as a whole. Indeed, if the collected data \eqref{eq:data} does not follow the patterns similar to what is shown in Fig.~\ref{fig:data_plot_chi_vs_T}, it has to be interpreted as a warning that some of the founding assumptions of spectroscopy were not upheld. For example, if the linear trend does not emerge, even for a very long duration or spectrometer length, it would mean that the noise field is not stationary or is not uniform, and consequently, the spectral density $S_\mathbf{n}(k,\omega)$ does not even exist. Only if the linear fits are possible to be performed with satisfactory precision, one can have a significant degree of confidence that the reconstructed spectrum is an accurate representation of the real thing.

\subsection{The \'{A}lvarez-Suter method for spectral density reconstruction}\label{sec:AS_method}

The final step of the reconstruction procedure is the deconvolution of the values of spectral density from the frequency and wavenumber combs found in spectroscopic formulas, or more precisely, in the slope \eqref{eq:slope}. 

This is achieved with the \'{A}lvarez-Suter method \cite{Alvarez_PRL11,Szankowski_JPCM17,Szankowski_PRA18} which takes as an input a dataset composed of slopes $A(k_p,\omega_p)$ generated for a choice of settings of $k_p$ and $\omega_p$  dictated by the shape of temporal and spatial combs. In particular, the required sets of $k_p$ and $\omega_p$ constitute of an arbitrary primary wavenumber $k_0$ and frequency $\omega_0$, that are then supplemented with auxiliary settings: $\omega_p = \omega_0,3\omega_0,\ldots,m_c\omega_0$, and for each multiple of primary frequency $m\omega_0$ a subset of wavenumbers, $k_p = m k_0, m(k_0+ k_d),\ldots, m(k_0+l_c k_d)$. The slopes produced in such a way can be arranged in a matrix form:
\begin{align}
\nonumber
&[A_{m,l}^{(k_0,\omega_0)}] =\\
\nonumber
&= [A\big( m(k_0+l k_d)\,,\,m\omega_0 \big) ]_{(m=1,3,\ldots,m_c;\,l=0,\ldots,l_c)} \\[.3cm]
\label{eq:recon_dataset}
&=\left[\begin{array}{lcl}
A\big(k_0\,,\,\omega_0\big)  & \ldots & A\big(k_0+l_c k_d\,,\, \omega_0\big)\\[.2cm]
A\big(3k_0\,,\,3\omega_0\big)  & \ldots & A\big(3(k_0+l_c k_d)\,,\, 3\omega_0\big)\\[.2cm]
\vdots & &\vdots\\
A\big(m_c k_0\,,\,m_c\omega_0\big)   & \ldots & A\big(m_c(k_0+l_c k_d)\,,\, m_c\omega_0\big)\\
\end{array}\right]\,.
\end{align}
By substituting the explicit form of Fourier coefficients \eqref{eq:c} into \eqref{eq:slope}, one obtains the following expression for the dataset slopes:
\begin{align}
\nonumber
&A_{m,l}^{(k_0,\omega_0)} =\sum_{m'=1,3,\ldots} \frac{4}{\pi^2 m'^2}\sum_{l'=-\infty}^\infty |\csp{l'k_d}|^2\\
\nonumber&\phantom{A_{m,l}^{(k_0,\omega_0)}=}\times S_\mathbf{n}\big(m' m(k_0+l k_d)-l' k_d\,,\,m' m\omega_0\big)\\
\nonumber
&\equiv \sum_{m''=1,3,\ldots}U_{m,m''}\sum_{l'=-\infty}^\infty |\csp{l'k_d}|^2\\
\nonumber
&\phantom{A_{m,l}^{(k_0,\omega_0)}=}\times S_\mathbf{n}\big(m''(k_0+l k_d)-l' k_d\,,\,m''\omega_0\big)\\
\nonumber
&\equiv \sum_{m''=1,3,\ldots}\!\!\! U_{m,m''}\!\!\!\sum_{l_{m''}=-\infty}^\infty\!\!\! V_{l,l_{m''}}^{(m'')}S_\mathbf{n}\big(m''k_0-l_{m''}k_d,m''\omega_0\big)\\
\label{eq:A=UAstar}
&\equiv \sum_{m''=1,3,\ldots}\!\!\! U_{m,m''}\!\!\!\sum_{l_{m''}=-\infty}^\infty \!\!\! V_{l,l_{m''}}^{(m'')} s^{(m'')}_{l_{m''}} \,\, ,
\end{align}
in which  the linear relation between the spectral density and the input slopes $A_{m,l}^{(k_0,\omega_0)}$ has been cast into vectorial form $\mathbf{A}^{(k_0,\omega_0)} = \mathbf{U}[(\mathbf{V}^{(m)}\mathbf{s}^{(m)})^\mathrm{T}]_{(m=1,3,\dots,\infty)}$, with matrix elements of $\mathbf{U}$ and $\mathbf{V}^{(m)}$ defined as
\begin{align}
\label{eq:U}
U_{m,m'} &= \sum_{n=1}^\infty \frac{4}{\pi^2}\frac{1}{n^2}\delta_{m n,m'},\\
\label{eq:V}
V_{l,l_{m}}^{(m)} &= \sum_{n=-\infty}^\infty |\csp{n k_d}|^2 \delta_{n-m l,l_{m}}=|\csp{(l_{m}+m l)k_d}|^2,
\end{align}
and $\delta_{i,j}$ is the Kronecker delta, that equals $1$ if $i=j$ and $0$ otherwise.

The objective is to invert the relation \eqref{eq:A=UAstar}, $\mathbf{s}^{(m)} =(\mathbf{V}^{(m)})^{-1}[ (\mathbf{U}^{-1}\mathbf{A}^{(k_0,\omega_0)})_{m,l}]_{(l=0,\ldots,l_c)}$, so that the vector of spectral density values for each $m$ is extracted from the input dataset. Strictly speaking such an operation is impossible to execute, because $\mathbf{U}$ and $\mathbf{V}^{(m)}$ are rectangular matrices with one finite and one infinite dimension, and as such are not invertible. Therefore, the only way to solve this problem is to adopt the approximation where the infinite sums in \eqref{eq:A=UAstar} are truncated in such a way that they can be rewritten in terms of finite square matrices $\tilde{\mathbf{U}}$ and $\tilde{\mathbf{V}}^{(m)}$, while the reminder is relatively small in comparison to the retained portion of the series. In the case of the temporal comb, for which the Fourier coefficients exhibit the power-law decay $|c_{m\omega_p}|^2\sim m^{-2}$, the optimal approximation for the sum over $m''$ is also the simplest---the cut-off at index $m_c$, so that $\tilde{\mathbf{U}} = [U_{m,m'}]_{(m=1,3\ldots,m_c;m'=1,3,\ldots,m_c)}$. Due to the choice of auxiliary settings of the frequencies $\omega_p$ the inverse matrix $\tilde{\mathbf{U}}^{-1}$ is guaranteed to exist and it can be calculated analytically \cite{Szankowski_PRA18}. On the other hand, the structure of spatial combs is qualitatively different; in contrast to the temporal comb, the spatial Fourier coefficients $|\csp{lk_d}|^2$ are non-monotonic functions of their indices. Indeed, since each coefficient is a combination of finite number of oscillating terms [see \eqref{eq:s}] they exhibit a beating pattern with possible periodic revivals. Consequently, there are no obvious general guidelines for which terms of the sums over $l_m$ should be retained. Let us parametrize the choice of indices corresponding to  the retained terms with an $l_c+1$ element set $\mathcal{L}_m =\{ l_0^{(m)},\ldots,l_{l_c}^{(m)}\}$, so that $\tilde{\mathbf{V}}^{(m)} = [V_{l,l'}^{(m)}]_{(l=0,\ldots,l_c;l'\in\mathcal{L}_m)}$. The simplest example is $\mathcal{L}_m = \{ -l_c/2,-l_c/2+1,\ldots,l_c/2\}$, which corresponds to approximating the spatial comb with a segment spanned on an interval $[k_0 - l_c k_d/2,k_0+l_c k_d/2]$. A more sophisticated example, where more emphasis is placed on the distribution of spatial coefficients $|\csp{lk_d}|^2$, would be to choose a set of indices of $l_c+1$ largest comb weights. The goal of this strategy would be to prevent the enhancement of the error due to overlap of the comb remainder and parts of spectral density in the case when it is difficult to estimate the positions and widths of spectral lines. Whether $\tilde{\mathbf{V}}^{(m)}$ for a given choice of $\mathcal{L}_m$ are also invertible depends on the Fourier coefficients $\csp{lk_d}$, that are in turn determined by the spatial distribution of the qubit block. Assuming that $(\tilde{\mathbf{V}}^{(m)})^{-1}$ does exist, at least for some of the indices $m$, the reconstruction of spectral density can be finalized:
\begin{align}
\nonumber
&S_\mathbf{n}^{(\mathrm{AS})}\big(mk_0-l k_d,m\omega_0\big)\equiv\\
\nonumber
&\sum_{m'=1,3,\ldots,m_c}\!\!\!\! (\tilde{\mathbf{U}}^{-1})_{m,m'}\!\!\sum_{l'=0}^{l_c} \big((\tilde{\mathbf{V}}^{(m)})^{-1}\big)_{l,l'}\,A_{m',l'}^{(k_0,\omega_0)}\\[.2cm]
\label{eq:reconstruction}
&\approx S_\mathbf{n}\big(mk_0-l k_d,m\omega_0\big),
\end{align}
for $l\in\mathcal{L}_m = \{l_{0}^{(m)},\ldots,l_{l_c}^{(m)}\}$, and the truncated matrices defined as
\begin{align}
\nonumber
&\tilde{\mathbf{U}}= [ U_{m,m'}]_{(m=1,3,\ldots,m_c;m'=1,3,\ldots,m_c)} \\
\label{eq:def_U_trunc}
&\phantom{\tilde{\mathbf{U}}}	= \left[\,\sum_{n=1}^\infty \frac{4}{\pi^2}\frac{1}{n^2}\delta_{m n,m'} \right]_{(m=1,3,\ldots,m_c;m'=1,3,\ldots,m_c)},\\[.3cm]
\nonumber
&\tilde{\mathbf{V}}^{(m)} = [ V^{(m)}_{l,l'} ]_{(l=0,\ldots,l_c; l'\in\mathcal{L}_m)} \\
\label{eq:def_V_trunc}
&\phantom{\tilde{\mathbf{V}}^{(m)}}	= [\, |\csp{(l'+m l)k_d}|^2\, ]_{(l=0,\ldots,l_c; l'\in\mathcal{L}_m)}.
\end{align}

\section{Example of spatiotemporal spectral density reconstruction}\label{sec:examples}

We illustrate the applicability of the proposed spatiotemporal noise spectroscopy method with the reconstruction of the spectral density following the procedure outlined in Sec.~\ref{sec:dist_proc}. For this purpose we choose a simple model in which the spatial and temporal fluctuations are independent,
\begin{equation}
C(\mathbf{r},t) = C_\mathrm{s}(\mathbf{r})C_\mathrm{t}(t)\, ,
\end{equation}
leading to factorization of the spectral density
\begin{equation}
\label{eq:model_S}
S_\mathbf{n}(k,\omega) = S_\mathrm{s}(k)S_\mathrm{t}(\omega)\,,
\end{equation}
into $S_\mathrm{s}(k) = \int_{-\infty}^\infty e^{-ikx}C_\mathrm{s}(x\mathbf{n})dx$ and $S_\mathrm{t}(\omega)=\int_{-\infty}^\infty e^{-i\omega t}C_\mathrm{t}(t)dt$. An example of a type of environment that would be characterized by this kind of spectrum is a collection of uniformly distributed, statistically independent and similar sources of noise that affect the qubits according to coupling law dependent on the relative position $\mu(\mathbf{r}_j-x_q \mathbf{n})$, where $\mathbf{r}_j$ is the position of $j$th source. Then the noise field is given by $\xi(\mathbf{r},t) = \sum_{j}\mu(\mathbf{r}_j-\mathbf{r})\eta_j(t)$, where the independence and similarity translates to statistical properties of the temporal fluctuations, $\overline{\eta_j(t)\eta_{j'}(t')}= \delta_{j,j'}C_\mathrm{t}(t-t')$. The auto-correlation function of such a noise field reads $\overline{\xi(\mathbf{r},t)\xi(\mathbf{r}',t')} = C_\mathrm{t}(t-t')\sum_{j}\mu(\mathbf{r}_j-\mathbf{r})\mu(\mathbf{r}_j-\mathbf{r}')$, with the spatial component determined by the coupling law and the source distribution $\rho_\mathrm{scr}(\mathbf{r}'')=\sum_j\delta(\mathbf{r}''-\mathbf{r}_j)$,
\begin{align}
\nonumber
&\sum_{j}\mu(\mathbf{r}_j-\mathbf{r})\mu(\mathbf{r}_j-\mathbf{r}')=\\
&=\int d\mathbf{r}''\rho_\mathrm{scr}(\mathbf{r}'')\mu(\mathbf{r}'')\mu(\mathbf{r}''-\mathbf{r}+\mathbf{r}')
	=C_\mathrm{s}(\mathbf{r}-\mathbf{r}'),
\end{align}
where the uniformity of the distribution has been invoked, $\rho_\mathrm{scr}(\mathbf{r}''+\mathbf{r})=\rho_\mathrm{scr}(\mathbf{r}'')$.

Furthermore, we choose for both components of spectral density identical line shapes, given by the Lorentzian distribution $S_0(z)=2/(1+z^2)$, 
\begin{align}
\label{eq:model_Ss}
S_\mathrm{s}(k)&=\nu_\mathrm{s}^2x_c\frac{S_0\big(x_c(k+k_s)\big)+S_0\big(x_c(k-k_s)\big)}{2}\\
S_\mathrm{t}(\omega)&=\nu_\mathrm{t}^2 t_c\frac{S_0\big(t_c(\omega+\omega_s)\big)+S_0\big(t_c(\omega-\omega_s)\big)}{2}.
\end{align}
Here $\pm k_s$ and $\pm\omega_s$ are the characteristic wavenumber and frequency of the spectrum (the positions of spectral lines and their mirror images, as required by the symmetry of spectral density), and $\nu_{\mathrm{s}/\mathrm{t}}^2$ are the maximal intensities of each component of the spectrum. The correlation length and the correlation time of the noise field are denoted by $x_c$ and $t_c$, respectively. This choice of spectral line shape is not necessarily representative of actual physical systems. Instead of realism, our approach was to provide an example that is both nontrivial (with the spectra being non-monotonous), and relatively effortless to follow.

Here, the slopes $A(k_p,\omega_p)$ constituting the dataset matrices $[A_{m,l}^{(k_0,\omega_0)}]$ [see \eqref{eq:recon_dataset}], that would be found with the linear fit method of Sec.~\ref{sec:lin_fit} in actual experimental settings, were instead simulated by a direct numerical evaluation of \eqref{eq:slope} with our model spectral density, $c_{m\omega_p}$ given by \eqref{eq:c}, and $v_{lk_d}$ defined by the chosen spatial distribution of the block of qubits $\rho_0(x)$.

\begin{figure}[tb]
\centering
\includegraphics[width=\columnwidth]{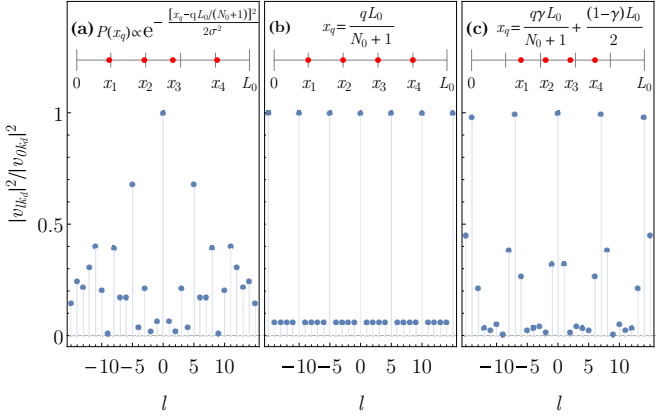}
\caption{The spatial Fourier coefficients $|\csp{lk_d}|^2$ for three cases of $N_0=4$ qubit block distributions $\rho_0(x)=\sum_{q=1}^{N_0}\delta(x-x_q)$ (illustrated by a schematic in the top part of each figure). (a) The irregular distribution used in our example, $x_q =\{ 0.19L_0,0.39L_0,0.56L_0,0.81L_0\}$. The qubit positions were chosen at random from Gaussian probability distributions $P(x_q)\propto \exp\{-[x_q-q L_0/(N_0+1)]^2/(2\sigma^2)\}$ with $\sigma=0.2 L_0/(N_0+1)$ (the last position was set to $x_{4}=L_0-x_1$). (b) The regular distribution with equidistant qubit positions $x_q = q L_0/(N_0+1)$. The coefficients form a repeating pattern with a period of $N_0+1$. (c) The ``compressed'' regular distribution with $x_q = q \gamma L_0/(N_0+1)+(1-\gamma)L_0/2$ and $\gamma=1/\sqrt{2}$. The coefficients exhibit revivals of the central peak with period $l_0=(N_0+1)/\gamma\approx 7.07$. Since $l_0$ is not an integer, the repeated peaks are slightly misaligned. }
\label{fig:spatial_F_coeffs}
\end{figure}

In our simulations we utilized the block comprised of $N_0=4$ qubits positioned at $x_q=\{ 0.19L_0,0.39L_0,0.56L_0,0.81L_0\}$. For this $\rho_0(x)$, the spatial Fourier series coefficients $|\csp{lk_d}|^2$ [see Fig.~\ref{fig:spatial_F_coeffs} (a)] are such that each $\tilde{\mathbf{V}}^{(m)}$ can be inverted for any choice of $\mathcal{L}_m$. Generally, the restrictions on the choice of $\mathcal{L}_m$ and the existence of $(\tilde{\mathbf{V}}^{(m)})^{-1}$ are encountered only for {\it regular} block distributions for which $x_q = q \gamma L_0/(N_0+1) + (1-\gamma)L_0/2$, where $0<\gamma\leqslant 1$. As it is illustrated in Fig.~\ref{fig:spatial_F_coeffs}, the Fourier coefficients of this type of distributions exhibit periodic revivals every $l_0\approx (N_0+1)/\gamma$. To illustrate why this can be a problem, let us consider a simple example of $\mathcal{L}_1 = \{-l_c/2,-l_c/2+1,\ldots,l_c/2\}$ (assuming $l_c$ is an even number). According to \eqref{eq:def_V_trunc}, if the dimension of $\tilde{\mathbf{V}}^{(1)}$ is greater than the revival period of Fourier coefficients (i.e., $l_c+1>l_0$), the $l$th and $(l_0+l)$th rows of the matrix  are identical (or are very similar, when $(N_0+1)/\gamma$ is not an integer). In that case, the matrix cannot be inverted because the determinant is zero (if rows are identical) or it is not well conditioned and an attempt at the numerical calculation of its reciprocal would be very inaccurate. To avoid such problems, the positions of qubits in the block used here were chosen at random from Gaussian distributions $P(x_{q<N_0}) \propto \exp\{-[x_q-q L_0/(N_0+1)]^2/(2\sigma^2)\}$ with $\sigma=0.2 L_0/(N_0+1)$, and $x_{N_0}$ set in such a way that $x_1+x_{N_0}=L_0$. The random shifts effectively ``deregularize'' the distribution, thus removing the revivals and periodicity. Instead, the Fourier series coefficients for large $l$ of such an irregular distribution scatter around average value $|\csp{0}|^2/N_0=N_0/L_0^2$ in a noisy fashion.

\begin{figure}[tb]
	\centering
	\includegraphics[width=\columnwidth]{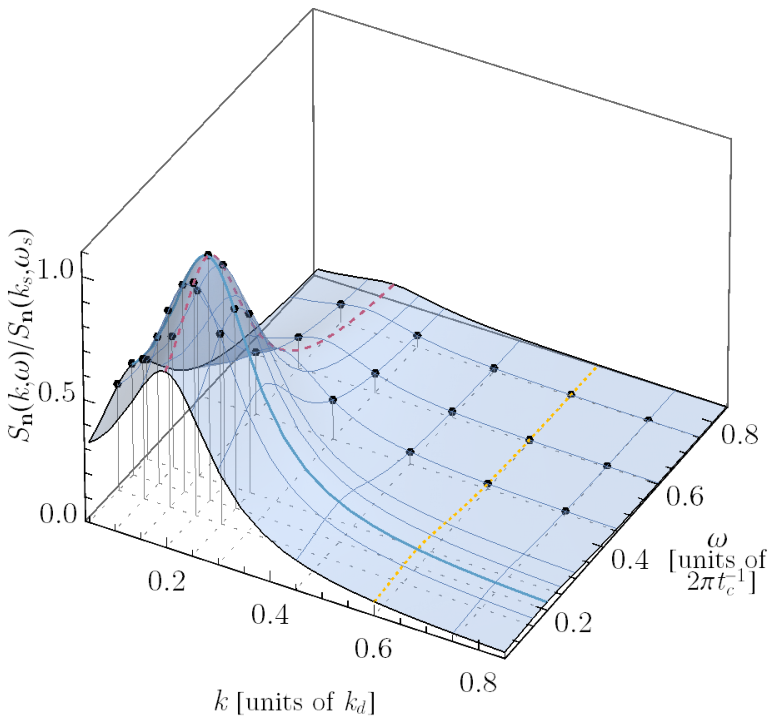}\vspace{.2cm}
	\includegraphics[width=\columnwidth]{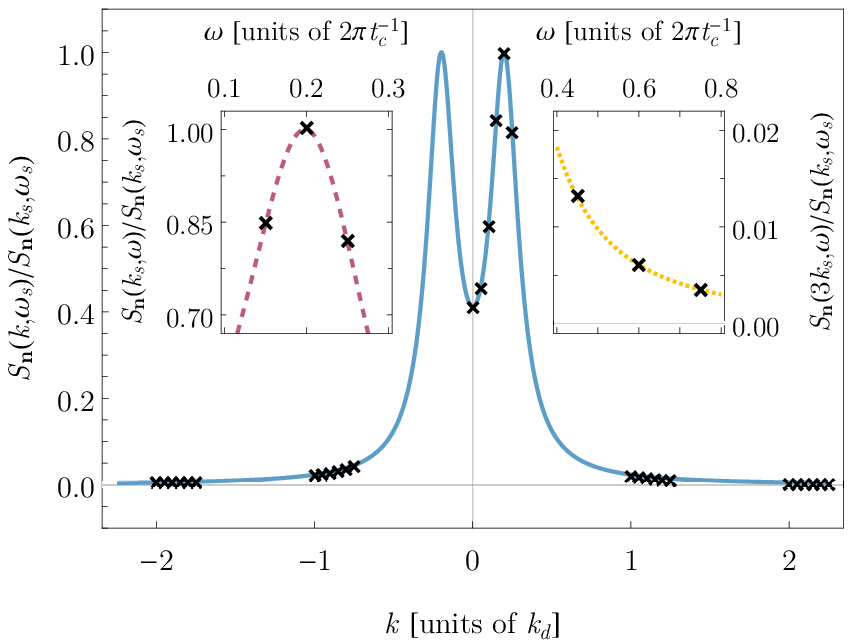}
	\caption{An example of spectral density reconstruction. (Top) The values found as a result of the procedure described in the text are compared with $S_\mathbf{n}(k,\omega)$ given by \eqref{eq:model_S}. (Bottom) The cross section along $k$ direction of the reconstructed spectral density. (Insets) The cross sections along $\omega$. The line colors and specifications (solid, dashed, etc.) correspond to the cross-sections highlighted in the top figure.}
	\label{fig:recon}
\end{figure}

We simulated the input matrices $\mathbf{A}^{(k_0,\omega_0)}$ for eighteen pairs of primary wavenumber and frequency settings combined from six choices of $k_0: \{0.00, 0.05k_d,0.10k_d,0.15k_d,0.20k_d,0.25k_d\}$ and three of $\omega_0 : \{0.15\times 2\pi/t_c,0.20\times 2\pi /t_c,0.25\times 2\pi/t_c\}$. The cut-off indexes were chosen to be $m_c=3$ and $\mathcal{L}_m = \{-2,-1,0,1,2\}$, so that each dataset matrix was $2\times 5$ dimensional. This means that, in total, we had to generate $10\times 18=180$ unique slopes. Then we computed the matrices $\tilde{\mathbf{U}}^{-1}$ and $(\tilde{\mathbf{V}}^{(m)})^{-1}$ (for $m=1,3$) according to Eqs.~\eqref{eq:def_U_trunc} and \eqref{eq:def_V_trunc}. Finally, we made use of the formula \eqref{eq:reconstruction} to obtain the values of the reconstructed spectral density, which are presented in Fig.~\ref{fig:recon}. In this example the reconstructed values of spectral density fit well the real course of $S_\mathbf{n}(k,\omega)$, for which we have chosen $x_c=1.5L_0$, $k_s=0.2k_d$ and $\omega_s=0.2\times 2\pi/t_c$.

\section{Conclusions}\label{sec:end}
We have demonstrated how a system of many qubits, arranged in a one-dimensional geometry, subject to pure dephasing due to environmental noise, and controlled with coordinated $\pi$ pulse sequences, can be used to reconstruct the spatiotemporal spectral density of the noise. The filtering of certain noise frequencies is achieved in a standard way---by using periodic pulse sequences. The key element of the scheme that turns the multi-qubit register into a spectrometer of spatial fluctuations, is the creation of a pattern of relative temporal shifts between $\pi$ pulse sequences applied to various qubits. When these shifts depend linearly on inter-qubit distance, the slope of the dependence together with the filter frequency of the sequences define the filter wavelength (equivalently a wavevector).    

Although the idea is rather simple and appealing, using such a spectrometer to obtain quantitative data on the spatiotemporal spectrum of the noise field requires more caution than when a single qubit is used to reconstruct the spectrum of temporal fluctuations of the local noise affecting it (note that the latter task, while routinely performed in recent years, is not entirely trivial either, especially in case of temporal spectra having peaks at finite frequencies  \cite{Szankowski_PRA18,Romach_arXiv18}). Large part of the paper has been devoted to detailed explanation of methods allowing for reliable extraction of ``spectroscopic'' information (spectroscopic formulas defined in Sec.~\ref{sec:spec_formula}) from raw measured data. The procedure of data analysis that we have discussed allows for self-diagnosis of the method, e.g.,~verifying if the noise is indeed stationary and spatially uniform, so that the spectrum can be in fact defined. It also allows for obtaining qualitative information about the noise---estimating correlation times and lengths, checking if the noise has a characteristic frequency or a wavevector---even before the spectral density is fully reconstructed. Further obstacles that have to be overcome are caused by the the fact that the spectroscopic information obtained
 from a single set of measurements involves multiple frequencies and wavevectors. This problem is known from single-qubit spectroscopy, in which the temporal spectrum at higher harmonics of the chosen filter frequency of the pulse sequence contributes to the measured signal. In the case considered here this issue becomes more significant due to the character of the wavevector filter generated in our scheme. We have given a thorough discussion of  modifications of the single-qubit \'{A}lvarez-Suter method necessary to deal with the case of multiple qubits.

The spatiotemporal noise field spectrum reconstructed with methods discussed here has a potential to become a very useful tool for characterizing physical properties of the complex system probed by the register. The main objection that could be raised against this point is the concern regarding the objectivity of the noise field, i.e., to what degree the presence of the register and its inevitable mutual interactions with the environment influences the statistics of the field it is used to probe. If this influence could not be neglected, the spectra reconstructed using variously configured registers would turn out to be manifestly distinct; in other words, different observers would report seeing different noise fields. In such an event, the noise could not be ascribed to the environment alone, and it could not be used as a reliable source of inference of environment's properties. One example of a system in which such a situation might occur are gate-controlled quantum dots hosting spin qubits \cite{Hanson_RMP07}. While the influence of charge noise on such qubits in well established \cite{Dial_PRL13,Kawakami_PNAS16,Yoneda_NN18}, the origin of this noise is a subject of discussion \cite{Ramon_PRB15,Beaudoin_PRB15}. Most importantly, it is not clear to what extent the noise is due to processes that depend on the spatial arrangement of metallic gates and voltages applied to them, which determine the positions of the quantum dots (and thus the qubits). In such systems the reconstructed spatiotemporal spectral density provides the same information as the complete set of temporal spectra of local noises and their cross-correlations; it can be used to characterize the decoherence suffered by the given register, which remains in fixed configuration ultimately designed for a different purpose then being a probe of the environment. On the other hand, the existence of objective noise field seems to be generally physically well motivated for systems where the influence of the environment on qubits can be described in the terms of coupling to fluctuating external electric \cite{Schindler_NJP13,Brownnutt_RMP15,Kumph_NJP16} and magnetic fields \cite{Taylor_NP08,Muhonen_NN14,vanderSar_NC15,Jakobi_NN17,Casola_NRM18}. A flagship example of such a system is a nitrogen vacancy center in a diamond acting as a sensor of fluctuating magnetic fields originating from {\it outside} of the diamond nanocrystal, e.g.,~generated by molecules \cite{Staudacher_Science13,Haberle_NN15,DeVience_NN15,Lovchinsky_Science16} and magnetic or superconducting materials \cite{Casola_NRM18} present in the vicinity of the piece of diamond in which the qubit is localized. What are the quantitative criteria for objectivity understood in this way remains an open question, obtaining the answer to which is one of the most vital issues for further studies on this subject. It is important to note that this question is not a novel one: It is equally applicable to the case of single-qubit spectroscopy, where the objectivity of temporal fluctuations affecting the qubit probe is also of concern. Even though there is no clear resolution to this problem, the success and practical importance of the dynamical-decoupling based spectroscopy method are undeniable. We believe that our multi-qubit extension will also prove to be viable despite these potential objections.
 
Aside from purely informative qualities of the noise field spectroscopy, the reconstructed spectrum $S_{\mathbf{n}}(k,\omega)$ could be utilized as an input for calculation of decoherence of different multi-qubit system arrangements and geometries, development of realistic quantum error correction protocols that take into account the presence of correlated errors caused by cross-correlations in noises experienced by the qubits, and development of optimized dynamical decoupling protocols, giving maximal enhancement of multi-qubit coherence for given set of practical constraints (number of pulses, minimal time delays between them etc.) \cite{Gordon_PRL08,Pasini_PRA10,Ajoy_PRA11,Wang_PRA13}. As a final note, we have focused here on the design of multi-qubit extension to the most commonly encountered single-qubit setting, where the noise is additive to the qubit energy splitting, and with the dynamical decoupling effected by short $\pi$ pulses. The analogous extensions to other kinds of single-qubit noise spectroscopy that involve transverse coupling to noise \cite{Cywinski_PRA14}, coupling to noise that is multiplicative to the control field used to rotate the qubit \cite{Frey_NC17,Norris_PRA18}, or using continuous driving of the qubit \cite{Willick_PRA18}, are consigned for future considerations.

\section*{Acknowledgements}
This work is supported by funds of Polish National Science Center (NCN), Grants no.~2015/19/B/ST3/03152 and no.~2012/07/B/ST3/03616

\appendix
\section{The Fourier transform of spatiotemporal filter function}\label{app:spf_ff}

First note that the time-domain filter function $f(t)$ defined in \eqref{eq:f_PDD}, can be decomposed into Fourier series \cite{Szankowski_PRA18,Szankowski_JPCM17}:
\begin{equation}
f(t) = \Theta(t)\sum_{m=-\infty}^\infty c_{m\omega_p}e^{im\omega_p t},
\end{equation}
where the Fourier series coefficients $c_{m\omega_p}$ are given by \eqref{eq:c}. We shall now use this form to calculate the temporal part of the Fourier transform \eqref{eq:spf_ft_def}:
\begin{align}
\nonumber
&\tilde{f}_{L; T}(k,\omega) = \int_{-\infty}^\infty dxdt e^{-ik x-i\omega t} f_{L;T}(x,t) \\
\nonumber
&= \int_0^T dt e^{-i\omega t}\int_0^L dx e^{-i kx}f\left(t+\tfrac{k_p}{\omega_p}x\right)\rho(x)\\
\nonumber
&=\sum_{m=-\infty}^\infty \!\!c_{m\omega_p}\int_0^T \!\!dt e^{i(m\omega_p-\omega) t}\int_0^L \!\!dx e^{i( m k_p-k) x}\rho(x)\\
\label{apdx:temp_Ft}
&=\sum_{m=-\infty}^\infty c_{m\omega_p}h_T(\omega-m\omega_p)\int_{0}^L dx e^{-i(k-mk_p)x}\rho(x)\,,
\end{align}
where the passband filter function $h$ is defined in \eqref{eq: sincs} and the qubit spatial distribution is given by \eqref{eq:periodic_rho}.

Since it is defined on a finite interval, the restricted qubit distribution can also be decomposed into Fourier series:
\begin{equation}
\Theta(L-x)\Theta(x)\rho(x) = \Theta(L-x)\Theta(x)\sum_{l=-\infty}^\infty \csp{l\frac{2\pi}{L}}e^{i l \frac{2\pi}{L} x},
\end{equation}
where the Fourier coefficients are given by
\begin{equation}
v_{l\frac{2\pi}{L}} = \frac{1}{L}\int_0^L e^{-i l \frac{2\pi}{L}x}\rho(x)dx=\frac{1}{L}\sum_{q=1}^N e^{-i l\frac{2\pi}{L}x_q}.
\end{equation}
Substituting this form into \eqref{apdx:temp_Ft} we get
\begin{align}
\nonumber
\tilde{f}_{L; T}(k,\omega) &=\sum_{m=-\infty}^\infty c_{m\omega_p}h_T(\omega-m\omega_p)\\
\nonumber
&\phantom{=}\times\sum_{l=-\infty}^\infty \csp{l \frac{2\pi}{L}}\int_0^L e^{-i\left(k-mk_p+l\frac{2\pi}{L}\right)x}dx \\
\nonumber
&=\sum_{m=-\infty}^\infty c_{m\omega_p}h_T(\omega-m\omega_p)\\
&\phantom{=}\times\sum_{l=-\infty}^\infty \csp{l \frac{2\pi}{L}}h_L\left(k-mk_p+l\frac{2\pi}{L}\right).
\end{align}
Now assume that $T=n_\mathrm{t}2\tau_p = n_\mathrm{t}T_0$ and $L=n_\mathrm{s}L_0$ where $L_0$ is the period of $\rho(x)$. The spatial Fourier coefficients calculated in respect to this spectrometer length satisfy the following relation (here $k_d = 2\pi/L_0$, so $2\pi/L = k_d/n_\mathrm{s}$):
\begin{align}
\nonumber
v_{l\frac{2\pi}{L}} &= \frac{k_d}{2\pi n_\mathrm{s}}\sum_{r=0}^{n_\mathrm{s}-1}\sum_{q=1}^{N_0}e^{-i l \frac{k_d}{n_\mathrm{s}}(x_q+r L_0)}\\
\nonumber
	&=\frac{k_d}{2\pi n_\mathrm{s}}\sum_{r=0}^{n_\mathrm{s}-1}\sum_{q=1}^{N_0}e^{-i l \frac{k_d}{n_\mathrm{s}}x_q}e^{-i l 2\pi\frac{r}{n_\mathrm{s}}}\\
\nonumber
	&=\left(\frac{1}{n_\mathrm{s}}\sum_{r=0}^{n_\mathrm{s}-1}e^{-i l 2\pi \frac{r}{n_\mathrm{s}}}\right)\frac{k_d}{2\pi}\sum_{q=1}^{N_0}e^{-i l\frac{k_d}{n_\mathrm{s}}x_q}\\
	&=\delta_{l,n_\mathrm{s}l'}\frac{k_d}{2\pi}\sum_{q=1}^{N_0}e^{-i l \frac{k_d}{n_\mathrm{s}}x_q}
	=\delta_{l,n_\mathrm{s}l'}v_{l' k_d},
\end{align}
where $v_{l'k_d}$ are identical to \eqref{eq:s}. Therefore, the spatial part of Fourier transform reads
\begin{align}
\nonumber
&\int_0^{n_\mathrm{s}L_0}\!\! dx e^{i(mk_p-k)x}\rho(x)=\\
\nonumber
&=\sum_{l=-\infty}^\infty v_{l\frac{2\pi}{L}}h_{n_\mathrm{s} L_0}\left(k-mk_p+l\tfrac{k_d}{n_\mathrm{s}}\right)\\
&=\sum_{l'=-\infty}^\infty v_{l'k_d}h_{n_\mathrm{s} L_0}\left(k-mk_p+l' k_d\right).
\end{align}


%

\end{document}